\documentclass[twocolumn]{emulateapj}

\shorttitle{WAT radio galaxy in the COSMOS field}
\shortauthors{V. Smol\v{c}i\'{c} et al.}

\def\comm#1   {{\tt (COMMENT: #1) }}
\def\new#1   {{\bf #1 }}
\def\etal{et al. }
\def\kms{~km~s$^{\mathrm{-1}}$}
\newcommand{\xmm} {{\sl XMM-Newton }}
\def\lss {LSS~\#17}
\slugcomment{  }

\begin{document}

\title{ A wide angle tail radio galaxy in the COSMOS field: 
        evidence for cluster formation\altaffilmark{0} }

\author{V.~Smol\v{c}i\'{c}\altaffilmark{1},
        E.~Schinnerer\altaffilmark{1}, 
        A.~Finoguenov\altaffilmark{2},     
        I.~Sakelliou\altaffilmark{1}, 
        C.~L.~Carilli\altaffilmark{3},
        C.~S.~Botzler\altaffilmark{4,5,6},
        M.~Brusa\altaffilmark{2},
        N.~Scoville\altaffilmark{7,8},
        M.~Ajiki      \altaffilmark{9},
        P.~Capak\altaffilmark{7},
        L.~Guzzo\altaffilmark{10},
        G.~Hasinger\altaffilmark{2},
        C.~Impey\altaffilmark{11},
        K.~Jahnke\altaffilmark{1},
        J.~S.~Kartaltepe\altaffilmark{12},
        H.~J.~McCracken\altaffilmark{13},
        B.~Mobasher\altaffilmark{14},
        T.~Murayama   \altaffilmark{9},
        S.~S.~Sasaki  \altaffilmark{9, 15},
        Y.~Shioya     \altaffilmark{15},
        Y.~Taniguchi\altaffilmark{15},
        J.~R.~Trump\altaffilmark{11} }
\altaffiltext{0}{Based on observations with the NASA/ESA {\em Hubble
Space Telescope}, obtained at the Space Telescope Science Institute, which
is operated by AURA Inc, under NASA contract NAS 5-26555; 
also based on observations obtained with XMM-Newton, an ESA science mission
with instruments and contributions directly funded by ESA Member States and
NASA; also based on data collected at: the Subaru Telescope, which is
operated by the National 
Astronomical Observatory of Japan; the European Southern Observatory, Chile;
Kitt Peak National Observatory, Cerro Tololo Inter-American Observatory, and
the National Optical Astronomy Observatory, which are operated by the
Association of Universities for Research in Astronomy, Inc.\ (AURA) under
cooperative agreement with the National Science Foundation; the National
Radio Astronomy Observatory which is a facility of the National Science
Foundation operated under cooperative agreement by Associated Universities,
Inc. Based also on observations obtained with MegaPrime/MegaCam, a joint
project of CFHT and CEA/DAPNIA, at the Canada-France-Hawaii Telescope (CFHT)
which is operated by the National Research Council (NRC) of Canada, the
Institut National des Science de l'Univers of the Centre National de la
Recherche Scientifique (CNRS) of France, and the University of
Hawaii. This work is based in part on data products produced at TERAPIX
and the Canadian Astronomy Data Centre.}
\altaffiltext{1}{ Max Planck Institut f\"ur Astronomie, K\"onigstuhl 17,
  Heidelberg, D-69117, Germany } 
\altaffiltext{2}{ Max Planck Institut f\"ur Extraterrestrische Physik,  
                  D-85478 Garching, Germany }
\altaffiltext{3}{ National Radio Astronomy Observatory, P.O. Box 0, 
                  Socorro, NM 87801-0387 }
\altaffiltext{4}{ University of Auckland, Private Bag 92019, Morrin Road, Glen
  Innes, Auckland, New Zealand } 
\altaffiltext{5}{ University of Canterbury, Private Bag 4800, Christchurch,
  New Zealand } 
\altaffiltext{6}{ Universit\"{a}ts-Sternwarte M\"{u}nchen, Scheinerstrasse 1,
  D-81679, M\"{u}nchen, Germany } 
\altaffiltext{7}{California Institute of Technology, MC 105-24, 1200 
                 East California Boulevard, Pasadena, CA 91125}
\altaffiltext{8}{Visiting Astronomer, Univ. Hawaii, 2680 Woodlawn Dr., 
                 Honolulu, HI, 96822} 
\altaffiltext{9}{Astronomical Institute, Graduate School of Science,
        Tohoku University, Aramaki, Aoba, Sendai 980-8578, Japan}
\altaffiltext{10}{Osservatorio Astronomico di Brera, via Brera, Milan, Italy}
\altaffiltext{11}{Steward Observatory, University of Arizona, 933 North Cherry
  Avenue, Tucson, AZ 85721}
\altaffiltext{12}{Institute for Astronomy, 2680 Woodlawn Dr., University of
  Hawaii, Honolulu, Hawaii, 96822} 
\altaffiltext{13}{Institut d'Astrophysique de Paris, UMR 7095, 98 bis
  Boulevard Arago, 75014 Paris, France} 
\altaffiltext{14}{Space Telescope Science Institute, 3700 SanMartin Drive,
  Baltimore, MD 21218} 
\altaffiltext{15}{Physics Department, Graduate School of Science
        and Engineering, Ehime University, 2-5 Bunkyo-cho,
        Matsuyama, Ehime 790-8577, Japan}

\begin{abstract}

  We have identified a complex galaxy cluster system in the COSMOS
 field via a wide angle tail (WAT) radio galaxy consistent with the idea
 that WAT galaxies can be used as tracers of clusters. 
The WAT galaxy, CWAT-01, is coincident with an elliptical galaxy 
resolved in the HST-ACS image. Using the COSMOS multiwavelength data set, 
we derive the radio properties of CWAT-01 and use the optical and 
X-ray data to investigate its host environment. The cluster hosting CWAT-01 
is part of a larger assembly consisting of a minimum of four X-ray luminous 
clusters within $\sim 2$~Mpc distance.
We apply hydrodynamical models that combine ram pressure and buoyancy forces
on CWAT-01. These models explain the shape of the radio jets only if the
galaxy's velocity relative to the intra-cluster medium (ICM) is in the range
 of 
about $300-550$\kms~which is higher than expected for brightest 
cluster galaxies (BCGs) in relaxed systems. This indicates that the 
CWAT-01 host cluster is not relaxed, but is possibly 
dynamically young. We argue that such a velocity could have been 
induced through subcluster merger within the CWAT-01 parent cluster and/or 
cluster-cluster interactions. Our results strongly indicate that we are 
witnessing the formation of a large cluster from an assembly of multiple 
clusters, consistent with the hierarchical scenario of structure formation.
We estimate the total mass of the final cluster to be approximately 
20\% of the mass of the Coma cluster. 
\end{abstract}


\section {Introduction}

Wide-angle tail (WAT) galaxies  form a class of radio galaxies, usually 
found in clusters,  whose radio jets have been bent into a 
wide {\bf C} shape. The general morphology of WATs suggests 
that the sources interact significantly with their external environment. 
The most natural interpretation of the jet bending is that the jets are 
being swept back by ram pressure resulting from the  high velocity motion of 
the associated active elliptical galaxy through its surrounding intra-cluster 
medium (ICM), first developed by Begelman, Rees \& Blandford (1979) and
applied by a number of investigators (e.g.\ Burns 1981, Pinkey \etal 1994).
In addition to ram-pressure, buoyancy forces were introduced
to explain the bending of the jets (e.g.\ Gull \& Northover 1973, Sakelliou
\etal 1996). If the jet density is lower than the density of the surrounding
medium, 
buoyancy forces will drag the jets towards regions of the ICM where 
the densities are equal. 

A point first noticed by Burns (1981) was that
WATs are usually associated with brightest cluster galaxies 
(BCG; D or cD galaxies), which are expected to reside at rest 
at the bottom of the clusters' gravitational potential well 
(Quintana \& Lawrie 1982, Merritt 1984, Bird 1994). 
Thus the large velocities of the WAT host galaxies relative to the ICM 
needed for the ram pressure models to shape the jets seemed to be inconsistent
with velocities typical for BCGs.\footnote{ Malumuth \etal (1992) have
    shown that 
  the velocity dispersion of the cD population is 0.3 of the dispersion of the
  cluster population; Beers \etal (1995) have found a velocity difference
  between the peculiar velocity of the central galaxy and the mean of the rest
  of the cluster galaxies $\lesssim 150$~\kms; recently Oegerle \etal (2005), 
 analyzing 25 Abell clusters, showed that peculiar velocities of cD galaxies
 differ only by $\sim 160$\kms\ from the mean cluster velocities. }
Therefore, it was necessary to evoke alternative scenarios to
explain the bent shape of WAT galaxies (e.g.\ Eilek 1984). 
However, the most prominent explanation is that the jets are bent by
ram pressure. 
It has been suggested in numerous studies that the necessary ram pressure
may be provided during cluster mergers (e.g.\ Pinkey, Burns \& Hill 1994,
Loken \etal 1995, Gomez 1997, Sakelliou 1996, 2000). 
This merger scenario is consistent with cosmological models, 
such as the cold dark matter model (CDM), which propose
that the structure in the Universe is formed hierarchically with
large features forming from mergers of smaller groups. 
The cluster potential well deepens then through accretion of poor clusters,
dark matter and gas into more massive systems. The material is accreted
from supercluster filaments which connect clusters into the large-scale structure
of the Universe (Evrard 1990, Jing \etal 1995, Frenk \etal 1996, Burns \etal
2002). 

Based on ROSAT PSPC X-ray observations of a sample of 9 Abell clusters 
containing WAT galaxies, Gomez \etal  (1997) find evidence for statistically 
significant X-ray substructure in 90\% ($8$ out of $9$) of the clusters hosting WATs, 
as well as a strong correlation of the orientation of the jets and the 
direction of X-ray elongation within the core of the cluster. Combined with 
numerical hydro/N-body simulations their results are consistent with 
WAT clusters undergoing mergers with groups or subclusters of galaxies.
Sakelliou \& Merrifield (2000) show 
that WATs are not generally located at the centers of their host clusters
as defined by their X-ray emission. They also find that the 
orientation of the bent jets is found to be preferentially pointed directly
towards or away from the cluster center. Thus, if the morphology is due to 
ram pressure, WATs are then primarily on radial orbits through the cluster. 
These results are explained as a natural consequence of cluster mergers
creating WAT galaxies (for details see Sakelliou \& Merrifield 2000 and 
references therein). 
Blanton \etal (2001) present optical imaging and spectroscopic observations
of environments surrounding 10 bent radio sources. 
They find that the clusters display a range of line-of-sight velocity
dispersions, $\sigma_{||}$,  from about $300-1100$\kms. 
The upper limit of $\sigma_{||}$ suggests that the host clusters are
either massive clusters and/or merging systems with significant substructure.

Since WATs are usually found in cluster environments, they can be used as
an efficient tool for cluster search, especially for high-redshift clusters 
where we are biased by the dimming of galaxies in the optical
and the ICM in X-ray emission. 
This approach has been successfully tested by 
Blanton \etal  (2000, 2001, 2003) using the VLA FIRST survey
   (Faint Images of the Radio Sky at Twenty centimeters survey; 
   Becker, White \& Helfand 1995) to 
search for galaxy clusters via WAT galaxies. The highest redshift cluster
they have identified to date is at $z=0.96$ (Blanton \etal 2003).

In this paper we discuss the properties of a WAT radio galaxy  (hereafter
CWAT-01) found in the VLA-COSMOS $2\Box^\circ$ survey, previously detected,
but not resolved by the  NVSS (NRAO VLA Sky Survey; Condon \etal 1998) survey
and not detected in the VLA FIRST survey
(Becker, White \& Helfand 1995).   
The multiwavelength data set of the COSMOS survey
(Scoville \etal 2006a) enables us to use the radio data to derive the
properties of this radio galaxy and the optical/X-ray data to investigate
its host environment.
In \S\ref{sec:data} we present the data utilized here. 
\S\ref{sec:wat} describes the radio and optical properties of CWAT-01.
In \S\ref{sec:cluster} we introduce the cluster and independently analyze its
properties in the X-ray and the optical. We discuss the results
in \S\ref{sec:discussion} and summarize them in \S\ref{sec:summary}.

For calculations in this paper, we assume
$H_0=70,\, \Omega_M=0.3, \Omega_\Lambda = 0.7$. 
We define the synchrotron spectrum as $F_{\nu} \varpropto \nu^{-\alpha}$, with
a positive spectral index, $\alpha > 0$, throughout the paper.

\section{ Observations and data reduction }
\label{sec:data}

\subsection{ Radio data }

The $2\, \Box^\circ$ COSMOS field was observed at $1.4$~GHz with the NRAO
Very Large Array (VLA) in A- and C- configuration for a total time of
$275$~hours (VLA-COSMOS survey;
Schinnerer \etal  2006).  The final  $2\, \Box^\circ$ map has a typical
  rms of $\sim10.5$~$(15)$~$\mathrm{\mu}$Jy/beam in the inner
  $1$~$(2)$~$\Box^\circ$ with a resolution of $1.5'' \times 1.4''$, thus making
  it the largest contiguous area covered in the radio-wavelengths regime with
  such a high sensitivity. The VLA-COSMOS large project catalog (presented in
  Schinnerer \etal 2006) contains $\sim 3600$ radio sources, $\sim90$ of which
  are clearly extended (most of them are double-lobed radio galaxies). The
  sensitivity of the survey combined with 
  the high resolution and the large area coverage makes the
  VLA-COSMOS project extremely valuable for e.g.\ studies of the sub-mJy
  radio population (i.e. the faint end of the radio luminosity function),
  dust-obscured star-formation history of the universe, evolution of
  radio-loud active-galactic nuclei (AGN). 
The survey utilized the 
standard VLA L-band continuum frequencies and the multi-channel continuum
mode. The complete data reduction was performed using the standard VLA
software AIPS. The A- and C-array data were combined in the $uv$ plane  
and then imaged using the task IMAGR. Cleaning boxes around bright sources
were defined manually, the two intermediate frequencies (IFs) and 
the left and right polarization were imaged separately and then combined 
into the final map. For more information about the survey and its scientific
objectives see Schinnerer \etal (2006). The local rms noise in the mosaic
around CWAT-01 is $\sim 10.5$ $\mathrm{\mu}$Jy/beam. 

Subsequent observations of CWAT-01
were obtained in June 2005 with the VLA in CnB configuration at 4.8 GHz.
The standard C-band continuum frequencies were used and the observations
were performed in the standard continuum mode. The CWAT-01 field 
was observed for 40 minutes on-source.
After flux and phase calibration  the data set was imaged using the 
AIPS task IMAGR. The IFs and the left and right polarization
were imaged together. Clean boxes were defined manually, the number of 
CLEAN iterations in
IMAGR was set to $3000$ and the flux cut-off to $60$ $\mathrm{\mu}$Jy. 
The $uv$ data points were weighted using the natural weighting function
($UVWTFN$ was set to 'N').
Due to the asymmetric $uv$ coverage the resolution of the 4.8 GHz map
was tapered down to obtain a rounder beam (the UVTAPER option 
in IMAGR was set to $35$ k$\mathrm{\lambda}$).
The resolution and  rms noise in the tapered map are
$7.27''\times 5.53''$ and $\sim 40$ $\mathrm{\mu}$Jy/beam, respectively.

\subsection{ X-ray data }
\label{sec:Xdata}

The COSMOS $2\Box^\circ$ field is being observed by the \xmm 
satellite (Jansen \etal 2001) for an awarded time of 
$1.4$~Msec (Hasinger \etal  2006). The data collected to date amount
to 0.8 Msec over the $2\Box^\circ$ area with an  effective depth of
$\sim40$~ksec, taking vignetting into account. 
In this study we utilize some of the results of 
Finoguenov \etal (2006) who identify clusters in the COSMOS field via diffuse
X-ray emission. The flux limit for the cluster identification in the 
$0.5-2$~keV energy band is $2\times 10^{-15}$~erg~cm$^{-2}$ s$^{-1}$.  
Finoguenov \etal (2006) report four diffuse structures within $4'$ of CWAT-01,
one of them containing CWAT-01.
We perform the spectral analysis of the extended emission
associated with the four identified clusters in the following 
way.
The EPIC pn observations of the COSMOS field have been merged
together and a uniform cleaning criterion for background flares has been
subsequently applied. Observations that do not satisfy this criterion
are removed. In this way we achieve a uniform background level for the
clean dataset. The 
resulting file consists of a total of $365$~ksec homogeneously 
cleaned data (for the whole $2\Box^\circ$ field). However, the region around
CWAT-01 is still covered at an effective depth of $43$~ksec as the removed
observations are all located at the edge of the COSMOS field. 

Since the instrumental background is not uniform over the detector (Lamb
  et 
al. 2002), in order to estimate the background, we produce a background file
from the same merged event file by excluding the area containing a
detected X-ray source.  This
removes $\sim 20\%$ of the area, which we account for in correcting for
the background. The background is further assumed to be the same in detector
coordinates. In calculating the background spectrum, the sky position of the
cluster is mapped to the detector, taking into account multiple pointings,
which map the same region of the sky on different detector areas. The
background spectrum is collected weighting accordingly the contribution in
each detector pixel. For the clusters in this study the ratio of the  
background to signal  accumulation times is 5,
which is sufficient to reduce the statistical uncertainty
associated with the background subtraction. The calculation of auxiliary
response files is performed by the SAS-based task {\it clarf} of Finoguenov
\etal (2004b),
which takes the mosaicing into account. The pn calibrations involved in the
data reduction correspond to SAS (Science Analysis Software; Watson \etal 2001)
version 6.5.

\subsection{ Optical data }

The optical imaging data of the $2 \Box^\circ$ COSMOS field, we use in this 
paper, was obtained in Spring 2004 and 2005. 
Within the COSMOS HST Treasury project the $2 \Box^\circ$ field was
imaged in 590 orbits during Cycles 12 and 13 using the 
Advanced Camera for Surveys (ACS; Scoville \etal 2006b). 
The F814W band imaging has a
$0.07''$ resolution and a $10\sigma$ sensitivity of
$I_{AB} = 27.2$.  Each of the 590 fields consists of 4 exposures, which are
calibrated and combined into a single image for each field using the
{\em MultiDrizzle} software (for details see Koekemoer \etal 2006.)
The whole COSMOS field was imaged with  
the Suprime-Cam camera (SUBARU telescope) in $6$ broad band filters, 
$B_J$, $V_J$, $g^+$, $r^+$, $i^+$, $z^+$ with
$5\sigma$ sensitivity in AB magnitudes of
$27.3$, $26.6$, $27.0$, $26.8$, $26.2$, $25.2$, respectively
(Taniguchi \etal 2006, Capak \etal 2006).
$u^*$ and  $i^*$ band images of the whole  $2 \Box^2$ COSMOS field
were obtained with the CFHT (Canada France Hawaii Telescope). The $5\sigma$
sensitivity in AB magnitudes is 
26.4 and 24.0 for $u^*$ and $i^*$, respectively.

The COSMOS photometric catalog was constructed using the SUBARU $i^+$ band
image. The details on constructing the 
photometric redshift catalog are described in Mobasher \etal  (2006).
The catalog produces photometric redshifts accurate to $dz/(1+z)=0.034$. 
 We utilize these redshifts to study the optical (sub)structure of the
  clusters in latter sections.

\subsection { Redshift }
\label{sec:redsft}

Trump \etal (2006) present spectroscopic redshifts of the first $\sim$500
X-ray and radio selected targets in the  $2\Box^\circ$ COSMOS field. 
The spectra were obtained using the Magellan IMACS instrument. They also
perform a robust classification of the observed objects  
(for details see Trump \etal 2006 and references therein). 
In the cluster area around CWAT-01 (see \S\ref{sec:cluster_ID})
there are two galaxies that have IMACS spectra.
Their properties, classification and redshifts are reported in 
Tab.~\ref{imacs_spec}. 

\begin{deluxetable*}{ccccc}
\tablecaption{\label{imacs_spec}}
\tablewidth{0pt}
\tablehead{
\colhead{ name } &
\colhead{ RA }  &
\colhead{ DEC } &
\colhead{ type } &
\colhead{ $z_{spec} \pm \Delta z_{spec}$ } }
\startdata
   COSMOS J100021.81+022328.5\tablenotemark{*} & 10 00 21.816 & +02 23 28.523
   & 
   elliptical & $0.22067 \pm 0.00007$ \\
   COSMOS J100025.30+022522.5 & 10 00 25.298 & +02 25 22.476  
   & elliptical & $0.22090
   \pm 0.00013$
\enddata
\tablenotetext{*}{Corresponds to SDSS J100021.81+022328.46 (see
   Tab.~\ref{sdss_spec})}
\tablecomments{Specifications of galaxies within the cluster area that 
have IMACS spectra (for details see Trump \etal 2006). The last column
   specifies the
spectroscopic redshift and error.}
\end{deluxetable*}

\begin{deluxetable*}{ccccccc}
\tablecaption{\label{sdss_spec}}
\tablewidth{0pt}
\tablehead{
\colhead{ name } &
\colhead{ plate } & 
\colhead{ fiber } &
\colhead{ RA }  &
\colhead{ DEC } &
\colhead{ $z_{spec} \pm \Delta z_{spec}$ } }
\startdata
  SDSS J10004.35+022550.71 & 501 & 348 &  10 00 4.354  & +02 25 50.711 &
  $0.2201 \pm 0.0001$ \\
 
  SDSS J10006.65+022225.98 &  501 & 353 &  10 00 6.654  & +02 22 25.982 &
  $0.2221 \pm 0.0002$ \\
 
  SDSS J100021.81+022328.46\tablenotemark{*} & 501 & 388 &  10 00 21.815 & 
       +02 23 28.463&  $0.2206 \pm 0.0001$ 
 
\enddata
\tablenotetext{*}{Corresponds to COSMOS J100021.81+022328.5 (see
  Tab.~\ref{imacs_spec})}
 
\tablecomments{SDSS specifications of galaxies within the cluster area that 
have SDSS spectra. The last column specifies the spectroscopic redshift and
error.}
\end{deluxetable*}

Searching for high resolution ($R=1800$) 
Sloan Digital Sky Survey (SDSS, York \etal  2000, 
Abazajian \etal 2003, 2004, 2005, Adelman-McCarthy \etal 2006) spectra, we
find that $3$ galaxies 
in our cluster area around CWAT-01 have SDSS spectra. One of them
was observed with IMACS. The specifications of these galaxies are 
listed in Table~\ref{sdss_spec}.
It is worth noting that the galaxies with spectroscopic redshifts are the 
most prominent galaxies within each cluster (see \S\ref{sec:cluster}).

The mean spectroscopic redshift of the galaxies presented above is $0.2209$ 
with an accuracy of $2.8\times 10^{-4}$. 
 Given the dispersion of the {\em photometric} redshift values they
are compatible with a redshift of $z=0.22$. 
Therefore, for the scope of this paper we adopt a mean cluster redshift
  of $z=0.22$, based on four measured {\em spectroscopic} redshifts.

\section{The wide angle tail galaxy: CWAT-01}
\label{sec:wat}

The radio galaxy first resolved in the VLA-COSMOS survey, CWAT-01, has a morphology 
typical for wide angle tail (WAT) galaxies. Its radio jets 
are bent into a wide {\bf C} shape (see Fig.~\ref{loop_1.4} for example).
In \S\ref{sec:radio_basics} we describe the structure of CWAT-01 and
derive its radio properties in
order to investigate the correlation to its host 
environment in latter sections. In \S\ref{sec:host} we describe the optical 
properties of the CWAT-01 host galaxy.

\subsection{ Radio properties of CWAT-01 }
\label{sec:radio_basics}

\subsubsection{The structure of CWAT-01}

The radio galaxy discussed here was first detected in the NVSS survey 
(NVSS J100027+022104; Condon \etal 1998), but not resolved due to the low 
resolution of the NVSS ($45$''). 
It was first resolved in the 1.4 GHz
mosaic of the central $1 \Box^\circ$ field from the VLA-COSMOS
pilot project (Schinnerer \etal 2004).
 CWAT-01 was not detected in the FIRST survey as this survey 
   over-resolves radio sources larger than $\sim10''$  (and thus underestimates
   their fluxes; see Becker, White \& Helfand 1995 for
   details). Hence, a galaxy like CWAT-01, which extends over more than $1'$
   on the plane of the sky and has a total flux density of
   $\sim13$~mJy (see below), would be strongly resolved out in the FIRST
   survey, i.e. the galaxy would consist of multiple components and the flux
   from the extended regions ($\gtrsim 10''$) of each component would be
   missed. The individual parts of 
   CWAT-01 that could have been detected by FIRST have flux densities below
   the detection limit  of the FIRST survey (1~mJy). Thus, CWAT-01 or
   any fraction of the galaxy stays undetected in the FIRST survey.

\begin{figure}
\includegraphics[bb = 14 14 237 210]{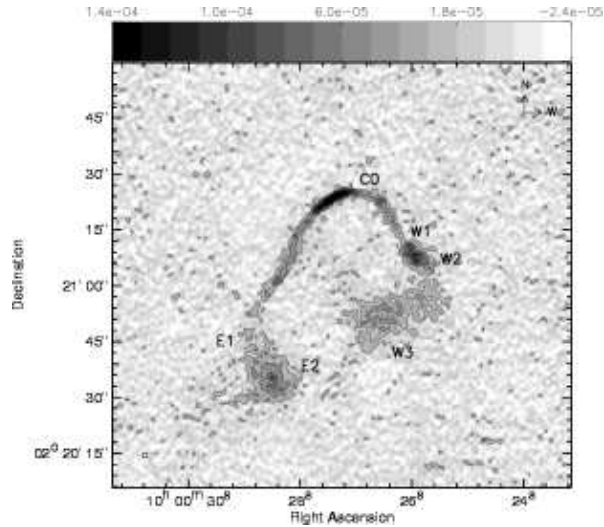}
\caption{$1.4$~GHz radio map of the wide angle tail 
galaxy CWAT-01 in the COSMOS field in grey scale with contours 
overlaid. The contour levels are in steps of $2\sigma$ starting
at the $2\sigma$ level ($1\sigma=10.5\, \mathrm{\mu Jy/beam}$). 
The clean beam is shown in the 
lower left corner (the resolution is $1.5'' \times 1.4''$).
$C0$, $E1-2$ and $W1-3$ label features of the jets 
discussed in the text. The colorbar units are in Jy/beam.
\label{loop_1.4}}
\end{figure}

Our new VLA-COSMOS observations, as part of the VLA-COSMOS large
project (Schinnerer \etal 2006), at 1.4 GHz provided 
significantly better data of this extended 
radio galaxy. The new 1.4 GHz VLA-COSMOS map of CWAT-01 is shown in 
Fig.~\ref{loop_1.4} (the resolution is $1.5''\times 1.4''$).
The radio jets of the galaxy are curved in a {\bf C} shape 
typical for WAT galaxies.  The central radio peak of CWAT-01 is 
at $\alpha=$~10~00~27.35   and $\delta=$~+02~21~24.15~(J2000) and its host 
galaxy is an elliptical galaxy with a photometric redshift of
$z=0.2\pm0.03$. The photometric redshift was taken from the
    COSMOS photometric redshift catalog described in detail in Mobasher \etal
    (2006). The quoted error is the $1\sigma$ error obtained from the 95\%
    confidence interval.   
The optical counterpart is discussed in more detail in  \S\ref{sec:host}. 
The jets are barely resolved in width (i.e. perpendicular to the 
jet axis) out to points E1 and W1 (see Fig.~\ref{loop_1.4}). 
The Eastern radio jet can be traced out to a projected distance of
$\sim 1'$ ($\sim 210$~kpc) from the central galaxy. 
At point E1 it bends to the west (in the projected plane), and broadens.
The end of the jet is marked with E2 (see Fig.~\ref{loop_1.4}).
The structure of the Western jet
is more complex:  It extends to a distance of
$\sim 45''$ ($\sim 160$~kpc) in the plane of the sky.
From the core till W1 it is narrow, but 
indicates curvature in the northern part.
After W1 the jet broadens and bends slightly to the west in the 
projected plane (W2). 
The faint feature labeled as W3 in Fig.~\ref{loop_1.4} seems also to be 
part of CWAT-01. 
The integrated flux density of CWAT-01 
at 1.4 GHz is $12.69$~mJy. This is within the errors of the reported 
NVSS flux density of $13.5\pm1.9$~mJy. For consistency 
we will use the flux density derived from the VLA-COSMOS for calculations 
throughout the paper.

The bending angle of the jets, which we define as the angle between lines 
 parallel to the part of the jet closest to the core 
(see Fig.~\ref{wat_cluster}, bottom panel), is $\sim 100^\circ$.
The asymmetry of the jets may be due
to projection effects which would indicate that the whole structure is
not moving only in the plane of the sky.  
However, with the data in hand we cannot reach any 
firm conclusions about projection effects nor rule them out. 
 More radio bands and higher resolution radio data where relativistic
 core beaming effects could be explored 
might resolve this issue; i.e. on $\sim10$~kpc scales from the
 radio core of the galaxy (where the bulk motion of the particles is
 relativistic) relativistic effects yield that the ratio
   of the radio brightness of the two jets can be correlated with the
 orientation angle.

\begin{figure}
\includegraphics[bb = 14 14 237 210]{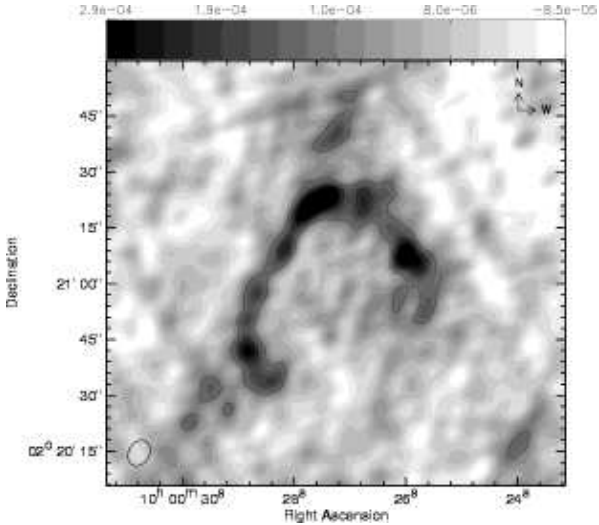}
\caption{$4.8$~GHz radio map of CWAT-01 with contours overlaid. 
The contour levels are in steps of $2\sigma$ starting
at the $3\sigma$ level ($1\sigma=40\, \mathrm{\mu Jy/beam}$). 
The clean beam is shown in the 
lower left corner (the resolution is $7.27'' \times 5.53''$).
The colorbar units are in Jy/beam.
\label{loop_4.8}}
\end{figure}

\subsubsection{ Spectral index }
\label{sec:sp_index}

\begin{figure}
\includegraphics[bb = 14 14 237 210]{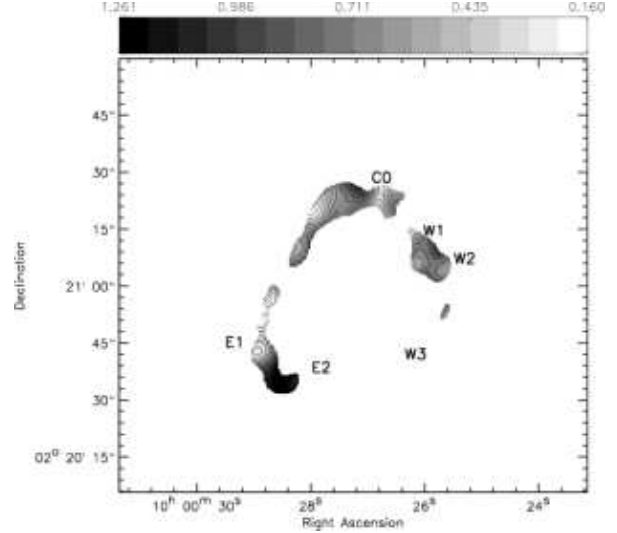}
\caption{Spectral index map of CWAT-01 for pixels with values 
$> 3\sigma$ in both maps ($1.4$~GHz and $4.8$~GHz).
C0, E1-2, W1-3 are labeled as in Fig.~\ref{loop_1.4} and 
are presented here for clarity.
We define the spectral index as $F_{\nu} \varpropto \nu^{-\alpha}$ 
throughout the paper. The contour levels are 0.1, 0.2, 0.3 etc.
\label{alpha}}
\end{figure}

The $4.8$ GHz map at $7.27''\times 5.53''$ resolution is shown in 
Fig.~\ref{loop_4.8}. 
The main features seen in the $1.4$ GHz map (with a resolution of 
$1.5''\times 1.4''$; Fig.~\ref{loop_1.4}) are still apparent although 
the much lower resolution reduces the amount of details.
The total integrated flux at 4.8 GHz is $4.1$ mJy. 
To obtain the  spectral index map (shown in Fig.~\ref{alpha}), 
the $1.4$ GHz image was convolved to the resolution of the $4.8$ GHz
map and the two images were regridded to the same pixel
scale.
Pixels with values below $3\sigma$ in each map were blanked.
The middle of the central feature is a flat spectrum region 
($0.1 \lesssim \alpha \lesssim 0.3$), with the spectral index steepening to
$\alpha \sim 1$ to the north-west edge in this feature. 
The outer region (corresponding to E1) with a flat spectrum 
($0.2 \lesssim \alpha \lesssim 0.6$)
suggests possible re-acceleration regions. The spectral index steepens to
$\alpha \sim 1$ at E2. In W2 the spectrum emission
is on average 
steeper than in the E1 feature, with an average 
spectral index corresponding to $\alpha \sim 0.7$.
The mean spectral index in the total source is $\alpha = 0.6$
which we will use for calculations throughout this paper, unless mentioned 
otherwise.

\subsubsection{ Radio power and luminosity }

We compute the radio power of CWAT-01 at 1.4 GHz using
a spectral index of $\alpha = 0.6$ (see \S\ref{sec:sp_index}) and the
luminosity distance ($1.1$~Gpc) 
at $z=0.22$. The radio power of CWAT-01 is then
$P_{1.4} = 2.0\times 10^{24}$ W Hz$^{-1}$ which places the radio
galaxy between FRIs and FRIIs where WATs are normally 
found (Hardcastle \& Sakelliou 2004). 

We also calculate the {\em total} radio luminosity, given by (e.g.\ O'Dea
  \& Owen 1987): 

\begin{eqnarray}
\label{lum1}
L_\mathrm{tot} = 1.2 \times 10^{27} D_{L,\mathrm{[Mpc]}}^2 F_0 \nu_0^{\alpha}
(1+z)^{-(1-\alpha)} \times \nonumber \\
    (1-\alpha)^{-1}(\nu_2^{1-\alpha} - \nu_1^{1-\alpha}) \,\, 
         [\mathrm{erg}\,\mathrm{s^{-1}}]
\end{eqnarray}

where $D_{L,[\mathrm{Mpc}]}$ is the luminosity distance expressed in Mpc 
and $F_0$ the flux density, at a fiducial frequency $\nu_0$, expressed in Jy.
We take 
the lower and upper frequencies to be $\nu_1 = 10$~MHz and
$\nu_2 = 100$~GHz, respectively. The observed frequency is
$ \nu_0 = 1.4$ GHz. The total luminosity of CWAT-01 is then
$ L_{tot} = 3.2 \times 10^{41}\, \mathrm{erg\,s^{-1}}$, typical for peculiar 
radio galaxies (Pacholczyk 1970).

\subsubsection{ Magnetic field and minimum pressure }
\label{sec:radio_pressure}

Assuming that the total energy in a radio galaxy is the sum of
the energy of electrons, $E_e$, heavy particles, $E_p$, and 
the magnetic field, $E_B$,
$ E_{tot} = E_e + E_p + E_B $, we can estimate the minimum energy
density, $u_{me}$, and the corresponding magnetic field, $B_{me}$, 
using the minimum energy condition which corresponds almost to
equipartition between the relativistic particles and the magnetic
field.\footnote{ Equipartition requires that the magnetic energy is
    equal to the total particle energy, i.e. $E_B=E_e + E_p$, while the
    minimum energy condition holds for $E_B=\frac{3}{4} \left(E_e +
      E_p\right)$. Hence, the computed total energy, $E_{tot}=E_e + E_p + E_B$,
    agrees within $\sim 10\%$ for the first and latter.  } 
We adopt the expression from Miley (1980; for details see
Pacholczyk 1970):
\begin{eqnarray}
\label{miley}
      u_{me} = \frac{7}{3}\frac{B_{me}^2}{8\pi} \,\, \mathrm{[dyn\, cm^{-2}]}
      \qquad \qquad  \qquad \qquad \qquad  \qquad \\
  B_{me} = 5.69\times 10^{-5}   \left [  \frac{1+k}{\eta}(1+z)^{3+\alpha} \times
               \qquad \qquad  \qquad \right. \nonumber \\
               \left. \frac{1}{ \Theta_x\Theta_y l
               \sin^{3/2}\Phi} \cdot 
               \frac{ F_0 }{ \nu_0^\alpha }\frac{\nu_2^{1/2-\alpha}
               - \nu_1^{1/2-\alpha}}{\frac{1}{2}-\alpha }  \right]^{2/7} \,\,
               \mathrm{[G]} \quad
\end{eqnarray}
where $k$ is the ratio of relativistic proton to relativistic electron energies, $
\eta$ is the  filling factor of the emitting region,
$z$ is the redshift, 
$\Theta_x$ and $\Theta_y$ correspond to the clean beam widths,
$l$ is the pathlength through the source along the line of sight, 
$\Phi$ is the angle between the  uniform magnetic field and the line of sight,
$F_0$ is the flux density at a fiducial frequency $\nu_0$,
$\nu_1$ and $\nu_2$ are the lower and upper frequency cutoffs and 
$\alpha$ is the spectral index. 
The minimum energy density and the corresponding magnetic field 
were measured in the middle of the diffuse portion of the Eastern radio jet
in the 1.4 GHz map with the following assumptions:
a) the radio plasma fills the volume completely ($\eta = 1$), 
b) the magnetic field is transverse to the line of sight ($\sin\Phi=1$), 
c) the relativistic proton energy equals the relativistic electron energy 
   ($k=1$),
d) there is cylindrical symmetry, and 
e) the radio spectrum spans  from $10$~MHz to $100$~GHz.
The mean spectral index derived from the spectral index map
in this part of the jet corresponds to  $\alpha = 0.65$. 
The resulting  magnetic field is $B_{me} = 3.7$~$\mathrm{\mu}$G and
the minimum energy density is $u_{me} = 1.3\times 10^{-12}$~$\mathrm{dyn\, cm^{-2}}$.
The minimum internal pressure within the jets is then 
$P_{min}=u_{me}/3 = 4.3\times 10^{-13}$~$\mathrm{dyn\, cm^{-2}}$.

\subsubsection { The particle lifetime }
\label{lifetime}

The synchrotron age of the electrons at frequency $\nu$ is 
given by van der Laan and Perola (1969) assuming the following model:
The electrons age as a result of synchrotron and inverse 
Compton losses due to the interaction with the cosmic microwave background 
(CMB).  There is a brief ``generation phase'', during 
which the relativistic gas is presumably created by the active galaxy, 
and a long-term ``remnant phase'' during which the particle supply is switched
off. The model computes the lifetime of the ``remnant phase'' as it assumes
that the lifetime of the ``generation phase'' is much shorter: 
\begin{equation}
\label{eq:t}
t \sim 2.6\times 10^4 \frac{B^{1/2}}{(B^2 + B_R^2){ [(1+z)\nu] }^{1/2}}
   \,\, [yr]
\end{equation}
where $B$ is the magnetic field in the jet and 
$B_R$ is the equivalent magnetic field of the CMB radiation, 
$B_R = 4[1+z]^4$ $\mathrm{\mu G}$. 
In order to constrain the electron lifetime, we substitute into 
eq.~[\ref{eq:t}] the magnetic field corresponding to the minimum energy
condition, $B=B_{me}$, calculated for the region at the end of the radio jets
at 1.4~GHz (E2 and W2 in Fig.~\ref{loop_1.4}). 
The 
magnetic field $B_{me}$ is again derived
making the same assumptions as in
\S\ref{sec:radio_pressure}. The mean spectral index at the end of the
Eastern and Western jet corresponds to $0.9$ and $0.7$, respectively,
and the minimum-energy magnetic fields are then $3.5 \, \mathrm{\mu G}$ and 
$3.1 \, \mathrm{\mu G}$, respectively. Both values give the lifetime of
an electron radiating at $1.4$ GHz of $\sim 13$~Myr.
Therefore, if we assume that there is no 
particle re-acceleration within the jets,
the relativistic electrons created in or near the core
could travel the whole jet length within their lifetime with
bulk velocities in the range of about $(0.04-0.05)c$.

\begin{figure}
\includegraphics[bb = -20 14 180 162]{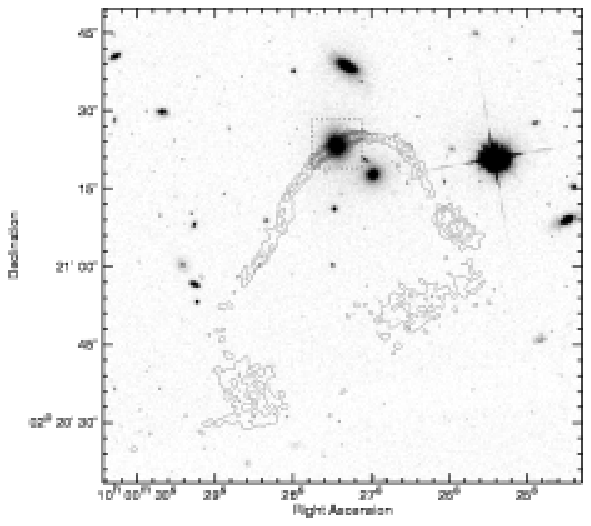}\\
\includegraphics[bb = -20 14 180 159]{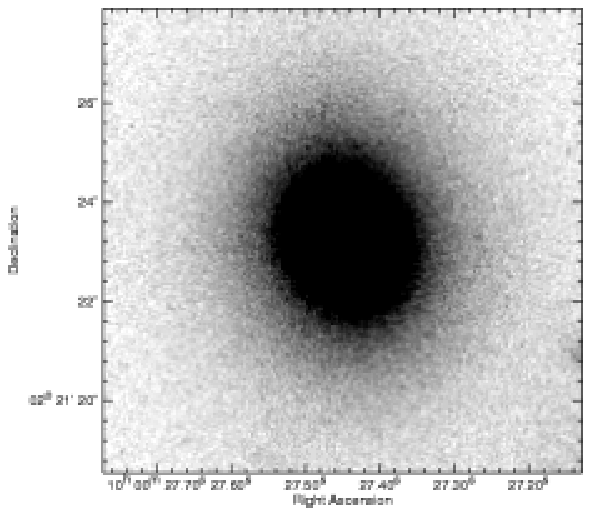}
\caption{
Top panel: HST-ACS FW814 band (grey scale) image of the CWAT-01 
   host galaxy with $1.4$~GHz radio contours    overlaid. 
   The contour levels are in steps
   of $2\sigma$ starting at the $3\sigma$ level 
   ($1\sigma=10.5\, \mathrm{\mu Jy/beam}$). 
   The dashed box indicates the region shown in the bottom panel.
Bottom panel:  HST-ACS FW814 band image of the CWAT-01 host galaxy.
   The galaxy has a morphology of an elliptical galaxy.
\label{host}}
\end{figure}

\subsection{ The host galaxy }
\label{sec:host}
\begin{figure}
\includegraphics[bb = 0 14 208 216]{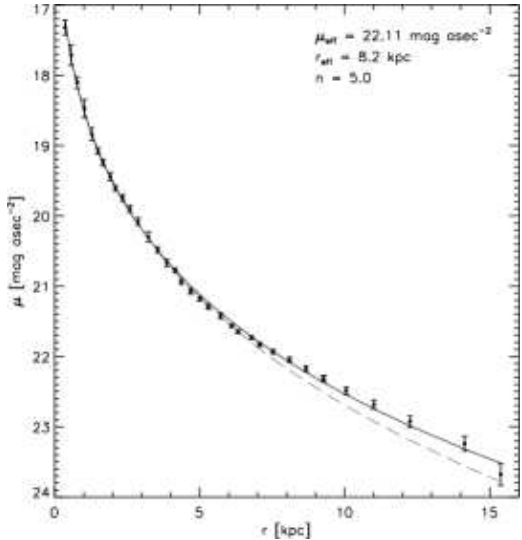}
\caption{ Surface brightness profile of the optical counterpart
of CWAT-01.
The surface brightness  is expressed in AB95 magnitudes 
and was derived from the HST-ACS F814W band image 
(see text for details). The dots 
are the data points with $1\sigma$ error-bars.
The de Vaucouleurs fit (dashed line) reveals an excess in surface
brightness in the outer parts of the galaxy while the Sersic law
fits the profile very well (solid line).
The effective surface brightness, effective radius and Sersic index 
for the redshift of $z=0.22$ are listed in the panel.
\label{loop_SB}}
\end{figure}

CWAT-01 is coincident with an elliptical galaxy, shown in Fig.~\ref{host}, 
located at $\alpha =$~10~00~27.43 and $\delta =$~+02~21~23.62~(J2000).
The spectral energy distribution (SED) type, reported in the COSMOS
photometric redshift 
catalog (Mobasher \etal  2006, Capak \etal  2006) is  
$1.33$ (ellipticals and  Sa/Sb correspond to 1 and 2, respectively; see 
Coleman \etal 1980 and  Kinney \etal  1996 for details).
The photometric redshift of the galaxy is $z=0.2\pm 0.03$ 
(the quoted error is the $1$~$\sigma$ error obtained from the 
95\% confidence interval).
We construct the surface brightness profile of the host galaxy
(shown in Fig.~\ref{loop_SB}) using the GIPSY ellipse fitting task 
ELLFIT on the background-subtracted HST-ACS F814W band image 
(Koekemoer \etal  2006). The surface brightness follows
the $r^{1/4}$ law fairly well, but it deviates from it
in the outer parts, indicating an excess in surface brightness, 
possibly an extended halo 
(see dashed line in Fig.~\ref{loop_SB}). The early type morphology, the
extensive envelope and the shallower surface brightness 
profile compared to the $r^{1/4}$ law suggest that the
CWAT-01 host galaxy might be classified as a D type galaxy 
(e.g.\ Beers \& Geller 1983).
To obtain a better fit to the surface brightness, we fit the Sersic model
$I(r) = I_{\mathrm{eff}} e^{ b_n [ 1 - (r/r_{\mathrm{eff}})^{1/n} ] }$, $b_n \sim 2n-0.324$,
(Sersic 1968) to the data with effective radius, $r_{\mathrm{eff}}$, 
effective intensity, $I_{\mathrm{eff}}$, and Sersic 
index, $n$, as free parameters (solid line in Fig.~\ref{loop_SB}). 
The data is very well fit by the Sersic law with 
$n=5.0$,  $r_{\mathrm{eff}} = 8.2$~kpc
and an effective surface brightness of 
$\mu_{\mathrm{eff}} = -2.5\log{I_{\mathrm{eff}}}-48.6= 22.1$~mag~arcsec$^{-2}$.
It has already been noted by Schombert \etal  (1987) that intrinsically
bright ellipticals are flatter (i.e. have higher values of $n$)  
and that intrinsically faint ellipticals have more curvature 
(i.e. lower values of $n$) than predicted by the $r^{1/4}$ law.
Typical values of $n$ for brightest cluster 
galaxies (BCGs) are $n>4$ (Graham \etal  1996). 
The effective radius and the Sersic index of the CWAT-01 host galaxy make
it consistent with being a BCG at the low end of the $n$ vs. $r_{\mathrm{eff}}$ 
correlation for BCGs (for details see Graham \etal  1996).
In \S\ref{sec:cluster_ID} we show that the CWAT-01 host galaxy is indeed
the brightest galaxy in its parent cluster.

\section{ The clusters }
\label{sec:cluster}
WATs are normally found in cluster environments and are in general 
associated with the brightest cluster galaxy (BCG). 
The X-ray image of the field around CWAT-01 (described in \S\ref{sec:Xdata}) 
showed the presence of multiple extended X-ray sources around the WAT. 
Additionally, in the previous section we have shown that the CWAT-01 
host galaxy has the characteristics of BCGs.
Taking advantage of the availability of the COSMOS multiwavelength data 
set we investigate the nature and properties of CWAT-01's environment in this
section. 

The whole cluster structure detected via diffuse X-ray emission
is part of the large-scale structure component, \lss, reported by 
Scoville \etal  (2006c). The diffuse X-ray emission shows
substructure itself. Throughout the paper we will use the
following nomenclature.
We refer to the whole area identified via diffuse X-ray 
emission as {\em cluster assembly}. The {\em cluster assembly} 
encompasses four subclumps
(i.e. four individual diffuse X-ray emitting regions), which we call
{\em clusters} or {\em poor clusters}, and 
is embedded in \lss; see for example Fig.~\ref{zones} (top panel).

\subsection{ X-ray properties }
\label{sec:cluster_x}

\begin{figure}
\includegraphics[bb = 0 14 216 166] {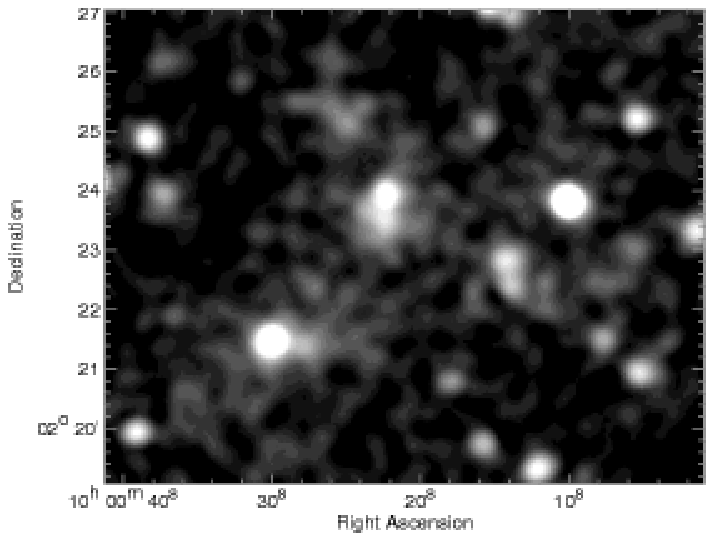}\\ 
\includegraphics[bb = 0 14 216 166]{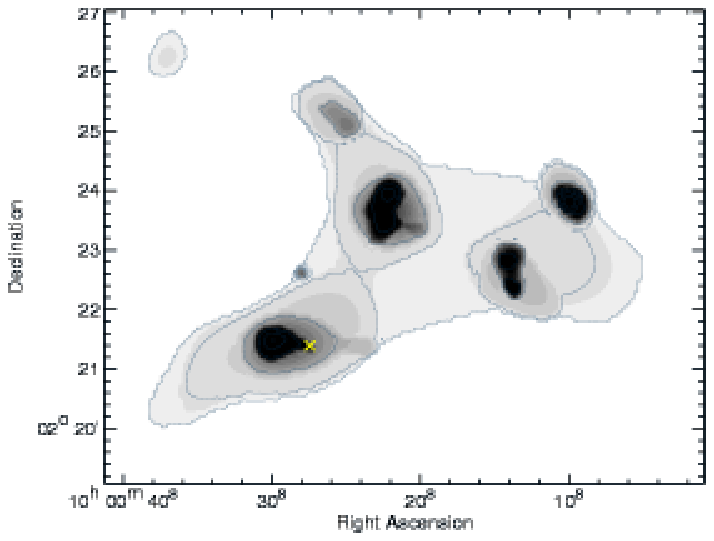}\\
\includegraphics[bb = 0 14 216 166]{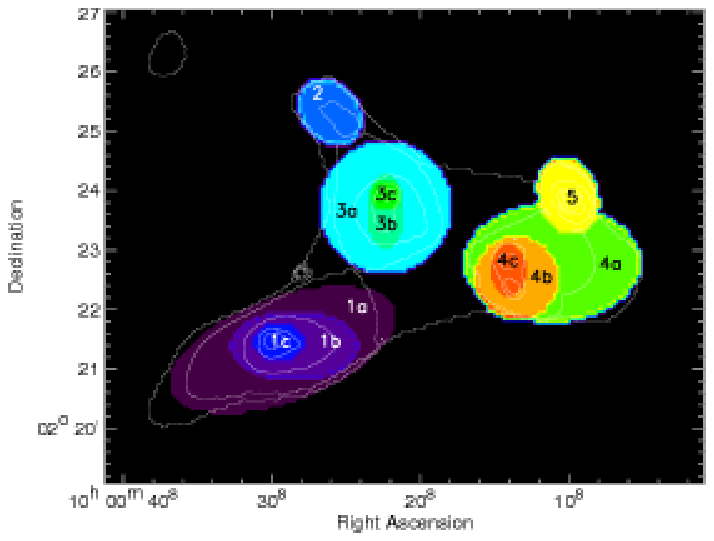}
\caption{
 Top panel: X-ray image of the cluster area in the $0.5-2$~keV energy
                band, convolved with a Gaussian of $8''$ width. The
                color-scale is linear with lighter color displaying higher
                signal to noise.
Middle panel: 
   Wavelet reconstruction of the X-ray $0.5-2$~keV band image showing
   the cluster area (taken from Finoguenov \etal 2006). 
   The contour levels are: 
   $3\times 10^{4}$, $10^{5}$, $3\times 10^{5}$, $10^{6}$, $3\times 10^{6}$,
   $5\times 10^{7}$~cnt/s/px. 
   The position of the CWAT-01 host galaxy is indicated (cross).
Bottom panel: 
   5 main zones with the corresponding subzones 
   constructed for the X-ray spectral analysis are shown in color
   (see also Tab.~\ref{t:x}).
   Each zone is labeled with its corresponding ID.
   X-ray contours (same as in top panel) are indicated to guide the eye.
\label{zones}}
\end{figure}
The search for extended X-ray sources in the COSMOS  $2\, \Box^\circ$
field (for details see Finoguenov \etal  2006) reveals 4 diffuse sources
   within $4'$ radius of CWAT-01.
 Fig.~\ref{zones} (top panel) shows a part of the X-ray image in the
  $0.5-2$~keV band encompassing the cluster assembly. In the middle panel of
  Fig.~\ref{zones} we display the same area in the sky, but from the wavelet
  reconstruction of the $0.5-2$~keV band image which was presented in
  Finoguenov \etal  (2006) and utilized for their cluster search. 
CWAT-01 is located in the south-eastern cluster (hereafter {\em CWAT-01 parent
cluster}).

Finoguenov \etal (2006) assign redshifts to the identified diffuse X-ray sources
by analyzing redshift bins of width $\Delta z = 0.2$ using the COSMOS photometric 
redshift catalog (Mobasher \etal  2006). 
Three of the four diffuse X-ray sources described
here are associated with a large galaxy concentration in the same redshift bin
with the median photometric redshifts of the clusters of $0.22$ 
(Finoguenov \etal  2006). A description of the cluster X-ray catalog names, 
their positions and fluxes is given in Table.~\ref{t:x0}.  
In the next sections we show that also the fourth diffuse X-ray source can be 
associated with an overdensity at a redshift of about $z=0.22$.

For the purpose of this paper we assume that 
the clusters are all located at the same redshift and calculate 
their properties at $z=0.22$.

\begin{deluxetable*}{ccccccc}[hb]
\tablecaption{\label{t:x0}}
\tablewidth{0pt}
\tablehead{
\colhead{ main } & 
\colhead{ catalog } & 
\colhead{ RA } &
\colhead{ DEC } &
\colhead{ Flux $\pm$ err } &
\colhead{ z$_{photo}$ }  \\
\colhead{ zone } & 
\colhead{ ID} & 
\colhead{  } &
\colhead{  } &
\colhead{ [$10^{-14}$ erg cm$^{-2}$ s$^{-1}$] } &
\colhead{  }
}
\startdata
1 & 78 &  10 00 28.337 & +02 21 21.6
   & 2.63 $\pm$ 0.15 & 0.22 \\
2 & 82 & 10 00 25.454 & +02 25 19.2
   & 0.76 $\pm$ 0.07 & 0.22 \\
3 & 85 & 10 00 21.852 & +02 23 42.0
   & 2.44 $\pm$ 0.15 & 0.22 \\
4 & 87\tablenotemark{*} & 10 00 13.925 & +02 22 48.0
   & 1.24 $\pm$ 0.10 & 0.40
\enddata
\tablecomments{
Description of the four X-ray luminous clusters adopted from 
Finoguenov \etal (2006): We list the main zone (column 1; see 
\S\ref{sec:x_spec} and Fig.~\ref{zones} for details) associated
with the X-ray cluster catalog ID (column 2), the cluster's position 
(columns 3 and 4), the corresponding flux and error (column 5) and 
the photometric redshift (last column). }
\tablenotetext{*}{Finoguenov \etal (2006) associate this extended X-ray
  emission with an optical overdensity at higher redshift. There seem to be,
  however, two X-ray features in their cluster catalog, one associated with
  this background cluster and the other with the overdensity we find at
  $z\sim0.22$. [Note, that the position of this cluster is not exactly
  coincident with the location of zone {\em 4a} (see Fig.~\ref{zones}),
  where we find the optical overdensity at $z\sim0.22$.] The revised cluster
  catalog (Finoguenov et al. in prep), produced utilizing the complete set of
  \xmm observations, will have a separate entry for the $z\sim0.22$ structure .  }   
\end{deluxetable*}

\subsubsection { The cluster assembly }
\label{sec:x_spec}

\begin{deluxetable*}{ccccc}
\tablecaption{\label{t:x}}
\tablewidth{0pt}
\tablehead{
\colhead{ zone } & 
\colhead{ kT [keV]} & 
\colhead{ normalization [$\times 10^{-5}$] } &
\colhead{ $\chi_r^2$ }  &
\colhead{ $N_{d.o.f.}$ }  }
\startdata
 1a & $1.08 \pm 0.26$ & $5.4 \pm 2.9$ & 1.06 & 33 \\
 1b & $2.26 \pm 0.74$ & $3.5 \pm 1.4$ & 1.12 & 19 \\
 1c & $2.37 \pm 0.45$ & $4.1 \pm 1.1$ & 2.07 & 16 \\
 2  & $1.40 \pm 0.45$ & $2.0 \pm 1.2$ & 1.38 & 16 \\
 3a  & $1.46 \pm 0.31$ & $4.5 \pm 2.2$ & 1.18 & 31 \\
 3b\tablenotemark{*}  & - & - & - & -\\
 3c\tablenotemark{*}  & - & - & - & -\\
 4a  & $1.47 \pm 0.61$ & $2.2 \pm 0.8$ & 0.98 & 19 \\
 4b\tablenotemark{*}  & - & - & - & -\\
 4c\tablenotemark{*}  & - & - & - & -\\
 5\tablenotemark{+}  & - & - & - & - \\
\enddata
\tablenotetext{*}{Nonthermal in origin}
\tablenotetext{+}{Foreground star}
\tablecomments{Results from the X-ray spectral analysis for the clusters
using the APEC thermal model. Column 1 displays the spectral 
extraction zone (see Fig.~\ref{zones} for reference). The temperature 
is given in column 2,
the normalization in column 3, the $\chi^2$ value in column 4, and the 
number of degrees of freedom, $N_{d.o.f.}$, in the last column.
}
\end{deluxetable*}

 For the spectral analysis we used the energy band of $0.5-3$~keV, since 
  the counts at energies above $3$~kev are dominated by background photons.
  Nonetheless, energies above $3$~keV can be used to check the quality of the
  background subtraction and this was found to be satisfactory. Based on the
surface brightness level, $5$ main zones were constructed for the
spectral analysis in such a way to avoid bright point-sources. The zones are
shown in Fig.~\ref{zones} (bottom panel). 
Zones {\em 1} (which corresponds to the  CWAT-01 parent cluster), 
{\em 3}, and {\em 4} are sampled with 3 subzones, labeled {\em a, b, c}. 
Table~\ref{t:x} summarizes the results of the spectral analysis
based on the APEC  thermal emission model (Smith \etal  2001). We report
the temperature, normalization and their corresponding errors, 
the reduced $\chi^2$ value and the number of degrees of freedom, $N_{d.o.f.}$. 
Zones {\em 3b, 3c}, {\em 4b}, and {\em 4c} turned out to be non-thermal 
in origin, likely AGN, while zone {\em 5} corresponds to a bright 
foreground star. 
 A power-law fit with a photon index of $\Gamma=2.3\pm 0.1$ to zone {\em
    1c} gives a better $\chi^2$ value ($\chi^2=1.2$ compared to $\chi^2=2.07$
  for the thermal model), which indicates that a more plausible interpretation
  of this zone may be a background AGN, as suggested by Brusa \etal
  (2006).  They associate
the optical counterpart of this X-ray peak with a
background ($z_{photo}=0.89$) galaxy. The galaxy has a morphology of 
a spiral galaxy, clearly resolved in the HST-ACS image. 
Nevertheless, we expect the X-ray centroid of the CWAT-01 parent
cluster to be located in the same area (i.e. associated with zone {\em 1c}).
The overall structure of the diffuse X-ray emission suggests that the center
is approximately in zone {\em 1c}. Furthermore, the center of mass calculated
using  stellar masses of the ``high-density'' galaxies (see
\S\ref{sec:cluster_opt} for details) is offset from the X-ray peak by $\sim
22''$ and is still within zone {\em 1c}. Thus, for simplicity we take the 
X-ray peak as the center of the CWAT-01 parent cluster for calculations in
this paper arguing that it is not far from where we would expect the cluster
center to be.  
In Tab.~\ref{t:x} we also present the estimated properties of 
zone {\em 1c}, assuming its thermal origin. The properties of this zone
do not deviate strongly from the expectations of cluster X-ray emission.
In addition, we emphasize that the mean values of the 
three-dimensional properties based on the spectral analysis results 
change only within $\sim 10\%$ when taking CWAT-01 as the cluster
center and are consistent within the errors 
with the properties calculated taking the X-ray peak as the center.

The temperature of each of the four clusters (see Tab.~\ref{t:x}) 
is consistent with the temperature range typical for poor clusters  
($1-3$~keV; Finoguenov \etal 2001).
Following Henry \etal  (2004) and Mahdavi \etal  (2005), we estimate the
cluster volume corresponding to the spectral extraction zones. The
derived values for gas mass, gas density, entropy and
pressure are listed in Tab.~\ref{t:x2}. 

\begin{deluxetable*}{cccccccc}
\tablecaption{\label{t:x2}}
\tablewidth{0pt}
\tablehead{
\colhead{ zone } & 
\colhead{ $ M_{gas} $ } & 
\colhead{ $ n_e  $  }  &
\colhead{ $ S $ } &
\colhead{ $ p $ } &
\colhead{ $ r_{min}, r_{max}  $} &
\colhead{ $ r_{500} $} &
\colhead{ $ M_{500}^{tot}  $} \\
                  &
\colhead{ $ [10^{11}\, M_\odot ]  $ } & 
\colhead{ $ [10^{-4}\,\mathrm{cm^{-3}}]  $  }  &
\colhead{ $ [\mathrm{keV\, cm^2}] $ } &
\colhead{ $ [10^{-13}\, \mathrm{dyn\, cm^{-2}}] $ } &
\colhead{ $ [\mathrm{Mpc}]  $} &
\colhead{ $ [\mathrm{Mpc}] $} &
\colhead{ $ [ 10^{13}\, M_\odot ]  $} 
}
\startdata
1a &
    13.0  $\pm$ 3.2 & 
    4.6 $\pm$ 1.1  &
    180.8  $\pm$ 52.8 &
    7.9 $\pm$ 2.7& 
    0.104, 0.460 & 0.490 & 5.8    \\
1b &
    5.9 $\pm$ 1.1  &
    6.7 $\pm$ 1.2  &
    295.6 $\pm$  103.4 &
    24.0 $\pm$ 9.0 &
    0.060, 0.315 & 0.490 & 5.8   \\
1c &
    2.00 $\pm$ 0.25  &
    23.7 $\pm$ 3.0  &
    133.4 $\pm$   27.5 &
    90.0 $\pm$  20.0 &
    0.000, 0.115 & - & - \\
\hline
2                   & 
    2.7 $\pm$ 0.7  &
    8.3 $\pm$ 2.3   &
    158.4  $\pm$  58.7 &
    18.8 $\pm$ 7.9 &
    0.000, 0.183 & 0.438 & 4.1   \\
3a                   & 
    8.5 $\pm$ 1.9   &
    6.0 $\pm$ 1.3   &
    204.2 $\pm$  52.3 &
    1.4 $\pm$ 0.4  &
    0.071, 0.253 & 0.449 & 4.4    \\
4a                   & 
    7.0 $\pm$ 1.2  &
    0.36 $\pm$ 0.06   &
    287.8 $\pm$ 121.6 &
    8.6 $\pm$ 3.8  &
    0.000, 0.375 & 0.451 & 4.4  
\enddata
\tablecomments{X-ray properties of the clusters. 
Column 1 lists the zones used for the spectral analysis (see
Fig.~\ref{zones}). The following columns are the gas mass
(2), gas density (3), entropy (4), pressure (5), 
the minimum and maximum radius of the extraction zone (6),
the radius corresponding to matter overdensity of 500 times the 
critical density, $r_{500}$, (7) and 
the total mass within $r_{500}$ (8).
}
\end{deluxetable*}

\begin{figure}
\includegraphics[bb = 14 14 126 119]{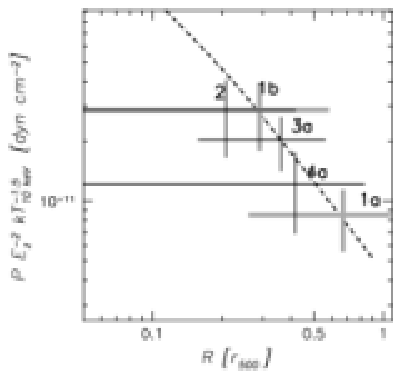}
\includegraphics[bb = 14 14 126 118]{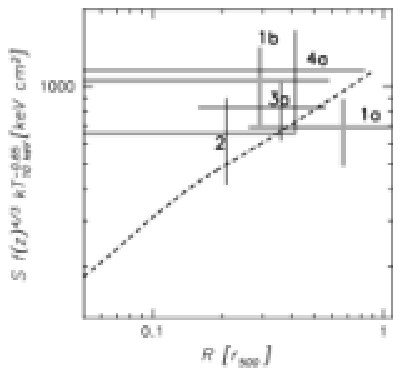}
\caption{ Pressure (left panel) and entropy (right panel) as a function of
  cluster radius for different spectral extraction zones (labeled in each
  panel, for comparison see Fig.~\ref{zones}). Thick crosses (zones {\em 1a,
  1b}) represent the CWAT-01 parent cluster. The length of the crosses
  indicates the $1\sigma$ errors. The dashed lines show the expected pressure
  and entropy  behavior based on local cluster studies and scaled to the
  redshift of this system (see text for details).
\label{entropy-pressure}}
\end{figure}

Using the luminosity-weighted temperatures of each of the four clusters 
(excluding zone {\em 1c}), we estimate the total mass within the 
corresponding radius of the enclosed matter overdensity of 500 times 
the critical density (M$_{500}^{tot}$, $r_{500}$ in the Table~\ref{t:x2}), 
using the $M$-$T$ relation from Finoguenov \etal  (2001). It is possible
that the masses are up to 20\% higher than the quoted values, according to
recent XMM and Chandra results on the $M$-$T$ relation (Arnaud \etal  2005;
Vikhlinin \etal  2005). The uncertainty in the total mass estimate is
primarily driven by the uncertainty in the measured temperature and is found
to be on the level of 40\% for the reported values.
In Fig.~\ref{entropy-pressure} we compare the derived properties of the X-ray
emission of the 4 clusters with the expectations based on 
local sample studies which we scale to the redshift of our cluster 
according to the expected evolution of shock heating 
(see Finoguenov \etal  2005 for details). The derived properties 
agree well with the prediction, which for shallow survey-type data, 
such as ours, is reassuring that the
identification, background subtraction and point source confusion issues have
been properly addressed.

\subsubsection {The CWAT-01 parent cluster }
\label{sec:wat_x}

The spectral properties of the CWAT-01 parent cluster are extracted from 
zones {\em 1a}, {\em 1b} and {\em 1c} (Fig.~\ref{zones}, bottom panel) 
as described in \S\ref{sec:x_spec}. 
The luminosity-weighted temperature of the CWAT-01 parent cluster 
(excluding zone {\em 1c}) is $\sim 1.7$~keV 
and the total mass $\sim 5.8 \times 10^{13} \, M_{\odot}$ which makes it
consistent with being a poor cluster (Finoguenov \etal 2001).
The spatial distribution of the diffuse X-ray emission of the 
cluster seems to be elongated and irregular.
In order to obtain an estimate of its spatial characteristics 
(i.e. the core radius, $r_c$, and the $\beta$ index), we 
obtain a 1-dimensional surface brightness profile  
using the $0.5-2.0$~keV background-subtracted 
image corrected for exposure time.
We fit the radial profile with a two-component model: 
   a) a Gaussian to describe the emission of the inner $\sim 20''$, and 
   b) a traditional $\beta$-model (Cavaliere $\&$ Fusco-Femiano 1976) for the
   underlying cluster.
The models are centered on the main peak of the X-ray emission.  A
$\beta$-model with $\beta = 0.57 \pm 0.06$ and 
$r_c = 48.0^{+8.7}_{-18.0}\,''=170.5^{+30.9}_{-63.9}\, kpc$ is a good
representation of the cluster X-ray emission.
The extended cluster component yields a count rate in
the (0.5-2.0)~keV energy range out to $r_{500}$ of 
$4.5 \times 10^{-3}$~cnt/s. We find that the luminosity of the cluster 
that hosts CWAT-01 is $3.6\times 10^{42}$~erg/s, 
consistent with L-T relation of Markevitch (1998), Mulchaey (2000), Osmond \&
Ponman (2004). 

Using the above derived
values of $\beta$ and $r_c$ and a temperature of $kT \sim 2.26$~keV (zone {\em
  1b}) we compute the central number density ($n_0$) as 
described in Sakelliou \etal  (1996).
The central number density corresponds to 
$n_0=1.085_{-0.08}^{+1.12} \times 10^{-3} \, \mathrm{cm^{-3}}$ which is
in agreement with the result of the spectral analysis (Tab.~\ref{t:x2}).
In \S\ref{sec:vgal} we use the derived quantities 
(i.e. $r_c$ and $\beta$) for hydrodynamical 
models explaining the bending of the jets of CWAT-01 and constraining 
the velocity of the host galaxy relative to the ICM.

\subsection{ Optical properties }
\label{sec:cluster_opt}

\subsubsection{ Cluster identification using overdensities: 
                Voronoi tessellations }

\label{sec:VTA}
To map the galaxy overdensity in the area of the X-ray cluster assembly
we use the Voronoi tessellation-based approach (hereafter VTA; e.g.\ Ramella \etal  2001, 
Kim \etal  2002, Botzler \etal  2006).
The VTA has several advantages over other overdensity estimators which make 
it favorable for the scope of this paper: 
First, no a-priori  assumptions  about cluster properties
     (e.g.\ density profile) are necessary making the technique 
     sensitive to elongated, i.e. non-symmetrical structures (Botzler \etal 2006). 
Secondly, we are mainly interested in substructure which can 
     efficiently be revealed with the VTA. 

A {\em Voronoi tessellation} on a two-dimensional 
distribution of points (called
nuclei) is a subdivision of the plane into polygonal regions (nuclei lying on the
edge may have open polygons), where each of them contains one and only one nucleus.
Each region consists of the set of points in the plane that are closer to that
nucleus than to any other. The algorithm used here for the construction of 
the Voronoi tessellation and calculation of the local densities 
is the ``varea'' code written by Botzler \etal (2006) 
which encompasses the ``triangle'' code by Shewchuk (1996).
Our goal is to quantify the clustering in the area where CWAT-01 is
located, thus the input for the Voronoi tessellation (nuclei) 
are galaxy positions drawn from the COSMOS photometric redshift catalog 
(Capak \etal 2006, Mobasher \etal  2006). 
The VTA then defines the effective area, $A$, that a galaxy 
occupies in the two dimensional space. 
Taking the inverse of the effective area gives the 
local density of the galaxy, $\rho_{local} = 1/A$.

The selection criteria we apply to the COSMOS 
photometric redshift catalog are the following: 
   we select objects classified as galaxies ($STAR\, < 1$)
   in the redshift bin of width $\Delta z = 0.2$ centered at the CWAT-01
   host galaxy's redshift reported in the catalog. 
The mean $2\sigma$ error of the selected galaxies is
$0.11 \pm 0.06$ in photometric redshifts. Thus, 67\% of our galaxies have
$2\sigma$ errors in photometric redshifts better 
than $0.17$ and 95\% better than $0.23$.

To robustly estimate the background density,
we apply the VTA to a region $\sim 10$ times larger than 
the region of interest which is $10' \times 8'$ corresponding to
$\sim (2.1 \times 1.7)$~Mpc$^2$ ($z=0.22$). 
In addition we run Monte Carlo simulations 
by randomly redistributing
the total number of galaxies in the analyzed field.
Then we apply the VTA
to each generated field and calculate the mean density, resulting in a
distribution of background densities with the corresponding standard deviation,
$\overline{ \rho_{bkg} } \pm \sigma_{bkg}$.
We define overdense regions as regions that have local density values of
$\rho_{local} > \overline{ \rho_{bkg} } + 10\sigma_{bkg}$.

\begin{figure}
\includegraphics[bb = -20 14 175 179]{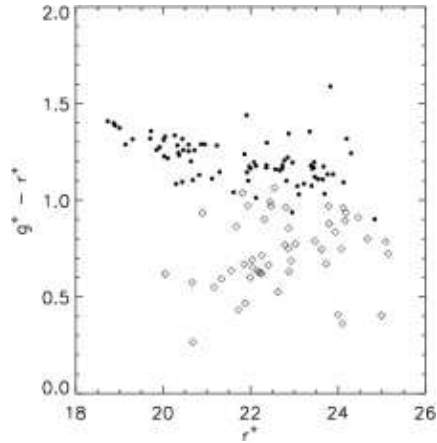}
\caption{ $g^+-r^+ \, \mathrm{vs.} \,r^+$ color magnitude diagram (CMD) 
using the COSMOS SUBARU $g^+$ and $r^+$ bands. The galaxies shown in the 
CMD are galaxies ($g^+<26.5$ and $Bj < 26.5$) within the 
cluster area of interest, corresponding to
$\sim (2.1 \times 1.7)$~Mpc$^2$ ($z=0.22$), which satisfy the 
overdensity criterion imposed in the VTA analysis:
$\rho_{local} > \overline{ \rho_{bkg} } + 10\sigma_{bkg}$ 
(see \S\ref{sec:VTA} for details). Galaxies in
masked-out regions (around saturated objects) are excluded. 
Filled symbols represent early type galaxies (SED type $< 2.5$) while
open symbols stand for late type galaxies (SED type $> 2.5$).
\label{cmd}}
\end{figure}
 Utilizing the COSMOS SUBARU $g^+$ and $r^+$ bands, we show in
  Fig.~\ref{cmd} the $g^+-r^+ \, \mathrm{vs.} \, r^+$ color-magnitude diagram
  of galaxies with local density values $\rho_{local} > \overline{ \rho_{bkg}
  } +  10\sigma_{bkg}$  and within the $\sim (2.1\times 1.7)$~Mpc$^2$ area
  encompassing the cluster assembly. 
Galaxies in masked-out regions (around saturated objects) 
are excluded to reduce the number of artifacts (note that excluding
the masked-out galaxies from the input sample for the VTA would
only slightly lower the mean background density value). 
We also impose a magnitude cut of $Bj<26.5$ and $g^+<26.5$
to exclude noise artifacts which are presumably due to the  
$g^+$ and $Bj$ detection limits for extended sources.
We therefore define the {\em final} sample of {\em ``high-density'' galaxies }
as galaxies that satisfy the following criteria: 
   a) $\rho_{local} > \overline{ \rho_{bkg} } + 10\sigma_{bkg}$,
   b) $Bj<26.5$ and $g^+<26.5$, and
   c) the galaxies are outside masked-out regions around saturated
      objects.

\subsubsection{ Cluster assembly structure }
\label{sec:cluster_ID}

\begin{figure}
\includegraphics[bb = 14 14 245 200]{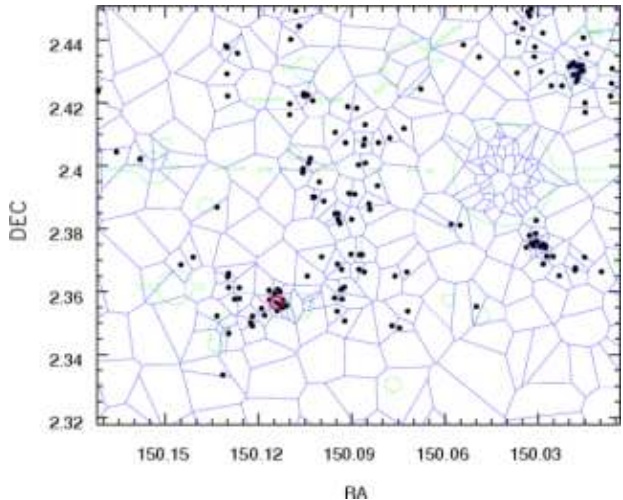}
\caption{ Voronoi tessellation analysis in the area of the cluster assembly  
(solid blue lines). The shown field is $\sim  (2.1 \times 1.7)$~Mpc$^2$ 
(at $z=0.22$) in size and $\approx 10$ times smaller than the total 
area analyzed. Masked-out regions (around saturated objects) in the 
photometric redshift catalog are marked with dotted green lines 
(see text for details).
The points represent ``high-density'' galaxies that  
   a) satisfy our overdensity criterion of 
      $\rho_{local} > \overline{ \rho_{bkg} } + 10\sigma_{bkg}$,
   b) satisfy the magnitude criterion of $g^+<26.5$ and $B_j<26.5$ and 
   c) are not located in masked-out regions. 
CWAT-01 is marked with the open solid circle. 
\label{vor_spider_net}}
\end{figure}

The Voronoi tessellation for the $\sim (2.1\times 1.7)$~Mpc$^2$ area 
surrounding the cluster assembly with indicated 
``high-density'' galaxies is shown in Fig.~\ref{vor_spider_net}.
The large-scale overdensity is elongated in NW-SE direction with
several obvious subclumps.
The spatial distribution of the galaxies seems not
to be spherically-symmetric, but irregular, both, on large and small
scales. In the areas around saturated stars (i.e. masked-out regions)
we loose all information about clustering.

\begin{figure*}[t]
\includegraphics[bb = 14 14 478 364]{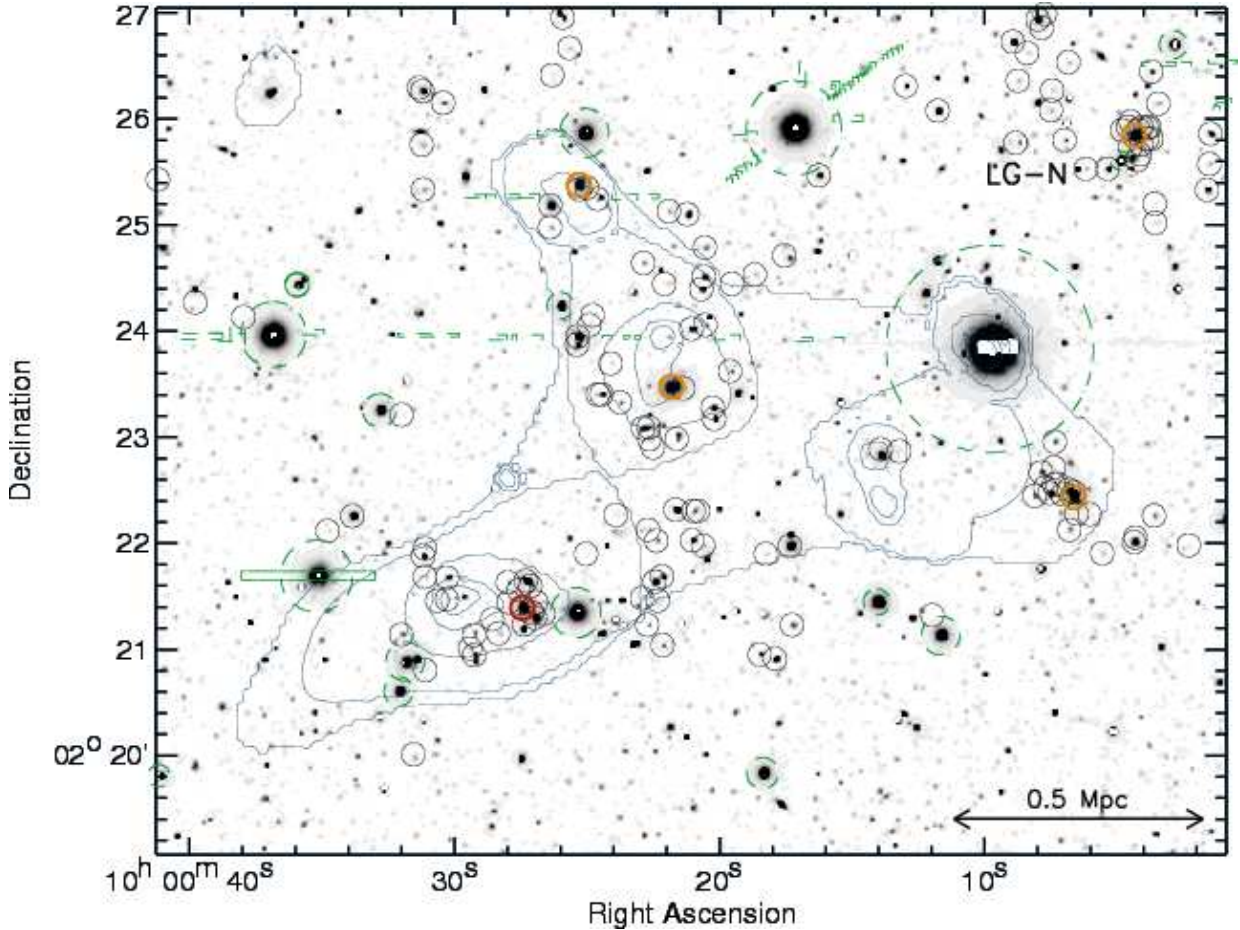}
\caption{Grey scale SUBARU $i^+$ band image of the cluster area  overlaid 
with X-ray contours (blue).  The contour levels are the same as in 
Fig.~\ref{zones}. The shown area is $\sim  (2.1 \times 1.7)$~Mpc$^2$ 
(at $z=0.22$) with thin circles (black) denoting the ``high-density''
galaxies (same galaxies as in Fig.~\ref{vor_spider_net}). 
Masked-out regions around saturated objects
(drawn from the COSMOS photometric catalog) are indicated with dashed green 
lines. CWAT-01 is marked with the thick red circle. Orange circles
indicate galaxies which have spectra (see \S\ref{sec:redsft} and
Tabs.~\ref{imacs_spec} and \ref{sdss_spec} for details).
LG-N labels the overdensity evident from the
Voronoi tessellation-based aproach (VTA) but not detected in X-rays 
(see text for details). The $0.5$~Mpc projected distance is 
indicated for reference.
\label{cluster}}
\end{figure*}
In Fig.~\ref{cluster} the ``high-density'' galaxies are overlaid
on the SUBARU $i^+$ band image.  For comparison, diffuse X-ray 
emission contours are also shown.
It is evident that the ``high-density'' galaxies display a 
complex and irregular spatial distribution, consistent with the
irregular and elongated distribution of the diffuse X-ray emission.
Each X-ray identified poor cluster has a counterpart in optical overdensities
approximately following the distribution of the X-ray emission.
Note that the X-ray cluster corresponding to zone {\em 4a} is
associated with an optical overdensity with a mean redshift
of $z\approx0.22$ like the other clusters. This is additionally
confirmed by the SDSS J10006.65+022225.98 galaxy spectrum 
(see Tab.~\ref{sdss_spec} for details).

The optical overdensities reveal, in addition, a clustering region
north-west from the diffuse X-ray emission (LG-N in Fig.~\ref{cluster})
not detected in the $0.5-2.0$~keV X-ray band.
The SDSS J10004.35+022550.71 galaxy spectrum confirms that this structure
is at the same redshift as the whole cluster system. 
We assume that LG-N is a loose group bound to the cluster assembly.

\subsubsection { Substructure in the CWAT-01 parent cluster }

\begin{figure}
\includegraphics[bb = 14 14 247 171]{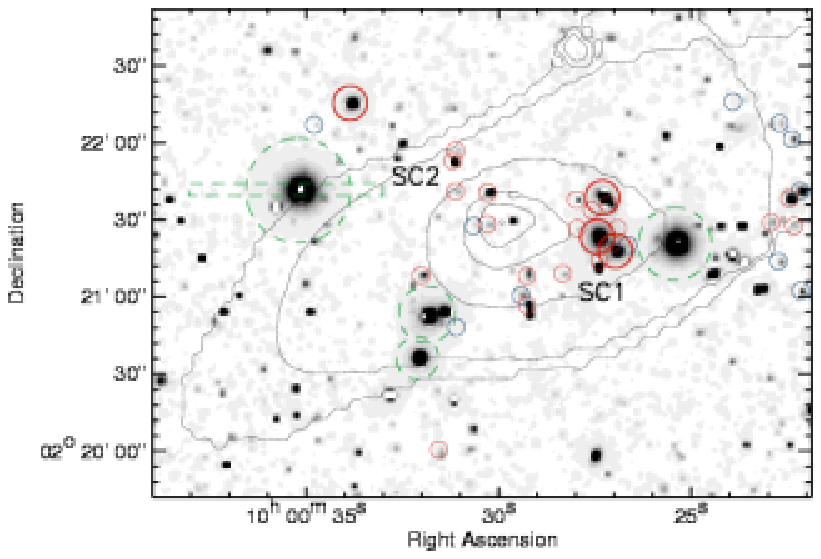}\\
\includegraphics[bb = 14 14 247 171]{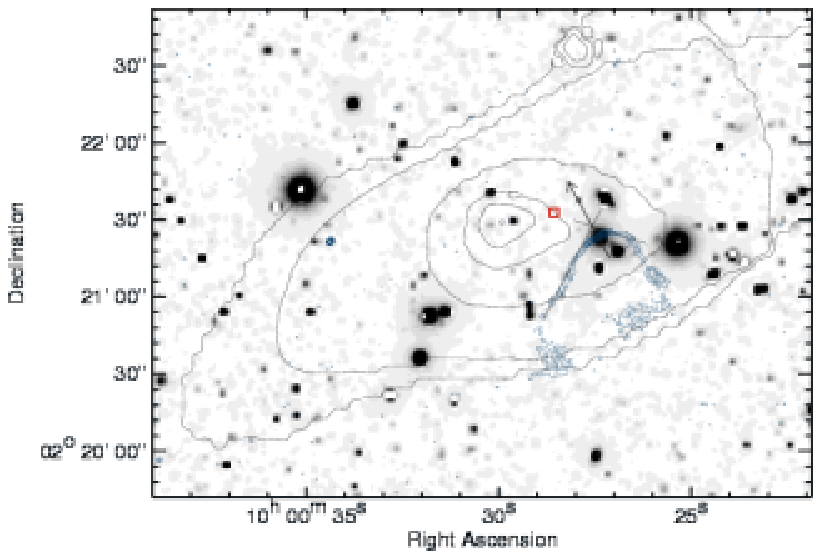}
\caption{
   Top panel: 
   SUBARU $i^+$ band image (grey scale) of the CWAT-01 parent cluster. 
   Overlaid are X-ray contours with contour levels as in 
   Fig.~\ref{zones}. Indicated are ``high-density'' galaxies.
   Red solid circles denote early type galaxies (SED type
   $< 2.5$) while blue solid circles indicate late type galaxies
   (SED type $> 2.5$). The brightest galaxies ($M_V < -20.5$) in 
   this area are marked with thick red circles. 
   Dashed green lines mark masked-out objects.  
   $SC1$ and $SC2$ label the two subclumps evident in the cluster 
   (see text for details). 
Bottom panel: 
   SUBARU $i^+$ band image (grey scale) and X-ray contours as in
   top panel overlaid with $1.4$~GHz radio contours. The radio
   contour levels start at the $3\sigma$ level and
   increase in steps of $1\sigma$.
   The bending angle of CWAT-01 is indicated by thin lines while 
   the arrow indicates the velocity direction of the galaxy.
   The open box marks the position of the center of mass of the cluster
   computed taking  into account the stellar masses of the 
   ``high-density'' galaxies in the CWAT-01 parent cluster (the stellar masses
   were drawn from the COSMOS photometric redshift catalog; Mobasher \etal
   2006). 
\label{wat_cluster}}
\end{figure}

The spatial distribution of the ``high-density'' galaxies 
(Fig.~\ref{wat_cluster}, top panel)
in the CWAT-01 parent cluster is irregular and elongated with two
dominant subclumps: 
   a) a Western overdensity (including the CWAT-01 host galaxy)
       extended in NW-SE direction ($SC1$ in Fig.~\ref{wat_cluster})
and
   b) an Eastern overdensity elongated in NE-SW direction ($SC2$).
There are three bright foreground stars contaminating the CWAT-01 parent
cluster area. Nevertheless, those masked-out regions should not affect our results
substantially since they are located at the outer parts of the parent 
cluster. 
Contrary to the expectation in relaxed systems where
one would expect early type galaxies 
to be centrally concentrated around the bottom of the cluster potential
well, the distribution of the early type galaxies (SED type $< 2.5$) 
in the CWAT-01 parent cluster is spatially elongated and coincident
with subclumps SC1 and SC2 (see Fig.~\ref{wat_cluster}).
Late type galaxies  are preferentially at the outskirts of the cluster.
The brightest galaxy in the cluster is the CWAT-01 host galaxy
($r^+=18.899\pm 0.004$, $M_V=-22.9 \pm 0.1$).
This is not surprising due to previous studies which have shown that 
WATs are generally associated with BCGs 
(e.g.\ Burns 1981). In addition,
the brightest galaxies ($M_V < -20.5$) in the cluster are strongly 
clustered in the region around CWAT-01, while only one (i.e. the second 
brightest in absolute V magnitude) 
is located at the outskirts of the cluster (see Fig.~\ref{wat_cluster},
top panel). 

Using the stellar masses reported in the COSMOS photometric redshift catalog
(Mobasher \etal 2006) of the ``high-density'' galaxies we compute the position
of the center of mass (indicated in Fig.~\ref{wat_cluster}, bottom
panel). The  offset of the center of mass from the main peak in the 
diffuse X-ray emission is only $\sim 22''$. 
Note that because of the bias introduced by the masked-out regions in the
cluster,  the center of mass may be closer to the main X-ray peak than 
given above.

\section{ Discussion }
\label{sec:discussion}

The unified theory for the mechanism responsible for bending the jets of WAT
radio galaxies is the dynamic pressure exerted on the jets by the ICM due to
the  relative motion between the galaxy and the ICM. 
In  \S\ref{sec:pressure} we compare the minimum pressure present in 
the radio jets to the thermal ICM pressure in order to investigate the
confinement of the jets.
We have shown in previous sections that CWAT-01 is associated with the 
BCG in its parent cluster. Therefore 
it is expected to be at rest in the minimum of the gravitational
potential (e.g.\ Bird 1994). In order to constrain the relative velocity
between CWAT-01's host galaxy and the ICM we apply several
hydrodynamical models explaining the bending of the jets of WATs
(\S\ref{sec:vgal}). 
In \S\ref{sec:merger} we suggest possible merger and encounter
scenarios responsible for the bending of the jets (\S\ref{sec:merger}).
The environment of CWAT-01 on larger scales (i.e. the cluster assembly) is
discussed in \S\ref{sec:assembly}.

\subsection{ Pressure balance }
\label{sec:pressure}
It is often assumed that the radio jets are confined by the ICM
(e.g.\ Miley 1980 and references therein) thus it is interesting to compare
the minimum internal pressure in the radio jets with the thermal ICM
pressure. The minimum internal pressure in the radio jets was calculated
in the middle part of the Eastern tail (\S\ref{sec:radio_pressure}),
$p_{min}^{jet} \cong 4.3\times 10^{-13} \mathrm{dyn\, cm^{-2}}$, and it
is lower than the ICM pressure in zone {\em 1b}, which
contains CWAT-01, 
$p_{ICM}^{zone\, 1b} = (24\pm 9) \times 10^{-13} \mathrm{dyn\, cm^{-2}}$
(see \S\ref{sec:cluster_x} for details). Such a pressure imbalance is not
unusual for WATs (e.g.\ Hardcastle \etal 2005). It implies either some
departure from the minimum energy condition (which is almost equal to
equipartition between the relativistic particles and the magnetic fields
in the jets) or a contribution to the pressure from particles which do not 
participate in equipartition such as thermal protons. However, one should  
be careful in comparing the two pressure values due to the low resolution of 
the  X-ray data and the numerous assumptions in the minimum pressure 
calculation. Nevertheless, the ram pressure models we apply to CWAT-01 in 
order to constrain the relative velocity between the galaxy and the ICM (next
section) should not be affected by this pressure imbalance. The inherent
assumption in these models is that the {\em dynamic} ram pressure of the ICM
is  comparable with the {\em pressure difference} across the jet.

\subsection{ Bending of the radio jets of CWAT-01: \\
             constraints on the galaxy velocity }
\label{sec:vgal}

In this section we apply several hydrodynamical models explaining
the bending of the jets of CWAT-01 in order to constrain the velocity of
CWAT-01's host galaxy relative to the ICM. The classical Euler equation
describes the jets if the bulk plasma velocity in the jets is
characterized by non-relativistic motions (e.g.\ Jaffe \& Perola 1973, 
Begelman \etal  1979,  Christiansen \etal  1981, Sakelliou \etal  1996).
Sakelliou \etal  (1996) developed a simple hydrodynamical model to
describe the bending of the jets of 3C34.16. They assume
that the forces acting on the jets are ram pressure and buoyancy and
they model the jets in the plane of the sky assuming
a steady plasma flow (Sakelliou \etal 1996, their equations [8]-[11]).
The strength of this model is that it provides a constraint on the
galaxy velocity relative to the ICM solely dependent on the jet density
at the point where ram pressure and buoyancy balance.  At the
turn-over point the bending of the jet changes its direction.
At this point the forces of ram pressure and buoyancy in the direction normal
to the jet balance and the only unknowns are the galaxy velocity and the
density of the jet at this point: 

\begin{equation}
\label{bend_point}
v_{gal}^2 = \frac{ 3\beta k T_{ICM} h }{ \mu m_p r_c }
            \frac { \frac{ r_{to} } {r_c} }{ 1 +  \left (\frac{ r_{to} }{ r_c }\right ) ^2}
            \frac{\bf \hat{r}\cdot \hat{n}}{{\bf \hat{v}_{gal}\cdot \hat{n}}}
            \left( 1 - \frac{\rho_j}{\rho_{ICM}}\right)
\end{equation}

Here $r_{to}$ is the radial distance from the cluster center to the turn-over
point (projected on the plane of the sky), 
$r_c$ and $\beta$ are the core radius and the standard 
hydrostatic-isothermal $\beta$ model parameter,
respectively (see \S\ref{sec:wat_x}),
${\bf \hat{n} }$ is the normal vector to the jet in the plane of the
sky, $v_{gal}$ is the component of the galaxy velocity
in the plane of the sky, $h$ is the scale height of the jet,
$\rho_j$ and $\rho_{ICM}$ are the
jet and ICM densities, respectively, $kT_{ICM}$ is the ICM temperature
in keV ($\sim 2.26$~keV), 
$\mu$ is the mean molecular weight and $m_p$ the proton mass.

\begin{figure}
\includegraphics[bb = 0 14 222 180]{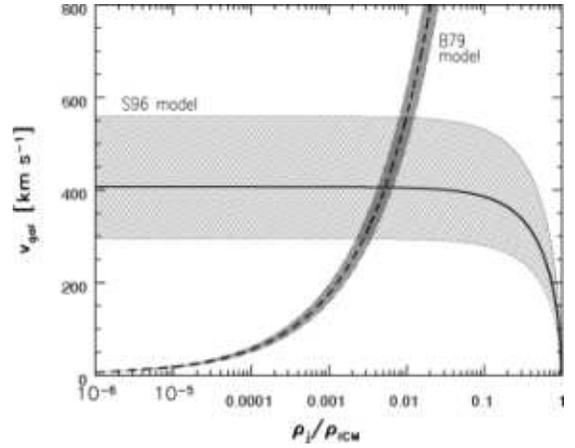}
\caption{
Mean galaxy velocity, $v_{gal}$, (thick lines) as a function of the ratio of  
the jet to ICM density, $\rho_j / \rho_{ICM}$, with
the allowed ranges (hatched regions) corresponding to our error estimates.
The broad hatched region shows the galaxy velocity calculated from
eq.~[\ref{bend_point}] using the model developed in Sakelliou \etal (1996;
S96) which takes into account ram pressure and buoyancy as forces responsible
for bending the jets (see text for details). The narrow hatched region shows
the galaxy velocity calculated from eq.~[\ref{begelman}] (developed byBegelman
\etal 1979; B79) taking only ram pressure into account. If we require that the
conditions of both models are satisfied, then the allowed ranges for $v_{gal}$
and $\rho_j$ are within the intersecting area. 
\label{vgal}}
\end{figure}

The 1.4 GHz radio map (Fig.~\ref{loop_1.4}) clearly shows that the radio jets
bend twice. Near the optical counterpart the jets turn to the
south. The second  bend of the jets is towards the  south-west direction
(points E1 and W1 in Fig.~\ref{loop_1.4}). The first bend can be attributed
to ram pressure as a result of the relative motion of the galaxy through 
the ICM. At larger radii buoyancy takes over and the jets are pushed towards
lower density regions in the ICM.
Fig.~\ref{vgal} shows the galaxy velocity as a function of the ratio
of the jet density to the ICM density, $\rho_j/\rho_{ICM}$, calculated
at the point W1 (thick solid line). 
The allowed range in velocities is indicated (broad hatched region).
The limiting velocity of the galaxy relative to the ICM 
(in the limit  $\rho_j/\rho_{ICM} \to 0$) is 
$v_{gal}\sim 400_{-110}^{+150}\,km\,s^{-1}$.
Note that $v_{gal}$ measures the projected velocity on the plane 
of the sky, thus any inclination of CWAT-01 to the line-of-sight 
would result in an additional line-of-sight component of velocity, 
thus increasing the total speed of the system. 
The upper and lower limits of the galaxy velocity were computed taking into
account the errors of $r_c$, $\beta$, and $kT$. 
Additional uncertainties are introduced through the 
estimate of ${\bf \hat{n} }$ at the turn-over point and  
${\bf \hat{r}\cdot \hat{n}}$ which depends on the position of the
center of the total mass of the cluster (discussed in \S\ref{sec:x_spec}).
Thus, the derived velocity is a rough estimate of CWAT-01's 
velocity relative to the ICM as a function of $\rho_j/\rho_{ICM}$.

Another hydrodynamical model, that we apply to  CWAT-01
to estimate the galaxy velocity, was first
proposed by Begelman \etal  (1979) to explain the jets in NGC1265. 
The curvature of the jets is again assumed to be produced by 
ram pressure exerted on the galaxy as it 
moves through the ICM. The ram pressure is
balanced by the centrifugal force exerted by the jet as it curves:

\begin{equation}
\label{begelman}
\frac{ \rho_{ICM} \, v_{gal}^2}{ h } = \frac{\rho_j \, v_j^2} {R}
\end{equation}

where $v_j$ is the bulk jet velocity,
$h$ is the scale hight,
$R$ is the radius of curvature, $v_{gal}$ is the
galaxy velocity and $\rho_{ICM}$ and $\rho_j$ are the ICM 
and jet densities, respectively.
Placing the mean jet velocity of the velocity range
derived in \S\ref{lifetime}, $v_j\sim 0.045c$, into 
eq.~[\ref{begelman}], using an estimate of the scale height 
(at point C0 in Fig.~\ref{loop_1.4}) 
of $h \sim 1.7''$ ($\sim 6$ kpc) and the radius of
curvature of $R\sim10''$ ($\sim 35$ kpc),  we show the galaxy 
velocity as a function of the $\rho_j / \rho_{ICM}$ ratio in 
Fig.~\ref{vgal} (dashed lines). Indicated is the galaxy velocity range 
corresponding to the jet velocity range, 
$0.04c \lesssim v_j \lesssim 0.05c$ (narrow hatched region).

Requiring that the conditions of both equations, 
eq.~[\ref{bend_point}]~and~[\ref{begelman}], are satisfied (as illustrated in 
Fig.~\ref{vgal}), we obtain an estimate of both, the galaxy velocity, 
$v_{gal}$, 
and the jet density, $\rho_j$. 
$v_{gal}$  is in the range of about 
$400^{+150}_{-100}\, km\,s^{-1}$   and  $\rho_j$ is
$0.005^{+0.01}_{-0.003}\, \rho_{ICM}$, respectively. The ICM density
is known from the spectral analysis, 
$\rho_{ICM} = (6.94\pm 1.27)\times10^{-4}\, cm^{-3}$, thus the estimated 
jet density with the corresponding errors is about
$0.03^{+0.09}_{-0.02} \times10^{-4}\, cm^{-3}$.

So far we have neglected possible {\em in situ} particle acceleration
within the jets. If particle re-acceleration occurs then 
the bulk lifetime of the synchrotron electrons in the 
jets would be higher than  the $13$~Myr estimated in 
\S\ref{lifetime}. A lower limit of the galaxy velocity can then be estimated
by assuming efficient conversion of kinetic energy into internal energy 
in the plasma jet flow (e.g.\ Eilek 1979).
If the observed luminosity, $L$, of the jets is supplied
by conversion of the bulk kinetic energy with an efficiency, $\epsilon$, then 
$L \sim \frac{\pi}{2} \rho_j h^2 v_j^3\epsilon$
(e.g.\ Burns 1981, O'Donoghue \etal  1993). 
Substituting $v_j$ in terms of the luminosity into 
eq.~[\ref{begelman}], one gets:

\begin{equation}
\label{v_min}
v_g \gtrsim \left(\frac{2L}{\pi} \right)^{1/3} \rho_j^{1/6} 
             \rho_{ICM}^{-1/2}
              h^{-1/6} R^{-1/2} \epsilon^{-1/3}
\end{equation}

which is only weakly dependent on the jet density, $\rho_j$. 
For an efficiency of $\sim 10\%$, assuming the above derived jet
density, the galaxy velocity is roughly $v_g \gtrsim 350$\kms~which is
consistent with the results from the previously applied models.  

The models we applied to CWAT-01 in the above discussion thus suggest 
that the galaxy velocity relative to the ICM is in
the range of about $300-550$\kms.

\subsection { Subcluster merging in the CWAT-01 parent cluster?}
\label{sec:merger}
Beers \etal  (1995) report a median velocity dispersion of 
$336\pm 40$\kms~  (which is in agreement with e.g.\  Ramella \etal 1994 and
Ledlow \etal 1996) 
for a sample of MKW/AWM poor clusters~\footnote{
   A sample of 23 poor clusters of galaxies originally identified by
   Morgan, Kayser \& White (1975) and Albert, White \& Morgan (1978).
}.
They find a velocity offset between the velocity of the central
galaxy and the mean velocity of the rest of the galaxies of 
$\lesssim 150$\kms~ for clusters with no evidence for subclustering. 
The velocity range of $300-550$\kms~
we found for CWAT-01's host galaxy, which is the dominant
galaxy (BCG) in its parent cluster, is significantly higher 
than this limit which  indicates recent merger events between less massive
systems of galaxies (Bird 1994).
High peculiar velocities are strongly correlated with the presence 
of substructure in the system (Bird 1994). 
Indeed, the VTA results indicate subclustering in 
the CWAT-01 parent cluster ($SC1$ and $SC2$ in 
Fig.~\ref{wat_cluster}). Furthermore, the brightest 
galaxies in the cluster are strongly concentrated around CWAT-01.
We speculate that SC1 and SC2 interact and, furthermore, that SC2 may
be infalling into the gravitational potential of SC1. This interaction
may cause a dynamical state of the cluster violent enough to produce the
inferred relative velocity of the CWAT-01 host galaxy to the ICM needed to bend
the jets in the observed way.
Moreover, the irregular assembly of early type galaxies in the 
CWAT-01 parent cluster (see Fig.~\ref{wat_cluster}) suggests that it 
is not a relaxed system. 
The elongated and irregular diffuse X-ray emission of
this cluster indicates independently possible merger or 
accretion events in the cluster. Finoguenov \etal (2005) have shown that 
X-ray elongations are often seen at the outskirts of clusters and that 
their spectral characteristics correspond to a colder, dense gas in pressure
equilibrium with the cluster. This gas is associated with accretion 
zones in clusters where dense parts of the filaments survive the
accretion shock and penetrate the cluster outskirts. 

Our results suggest that merger events within the CWAT-01 parent cluster
caused such a dynamical state in the cluster which is needed for bending the
radio jets of CWAT-01 in the observed way.
Such a merging scenario is 
consistent with conclusions of previous studies which have
suggested that the bent shape of WAT galaxies is caused by mergers 
(e.g.\ Gomez \etal  1997, Sakelliou \etal  1996, 2000).

 Based on a sample of $\sim20$ WAT galaxies which are located in Abell
  clusters and have ROSAT X-ray observations Sakelliou \etal (2000) have
  shown that WATs travel predominantly radially towards or away from the
  center (as defined by the X-ray centroid) of their host cluster. 
CWAT-01 does not seem to be on a radial motion in the 
projected plane of the sky, as is evident in Fig.~\ref{wat_cluster} 
(bottom panel). This may be a bias caused by projection effects. 
However, the gravitational influence of the
neighboring clusters may have played a significant role in causing the
inferred velocity modulus and direction of CWAT-01's host galaxy relative to 
the ICM.

Based on the above arguments we suggest that the radio jets
of CWAT-01 were bent as a consequence of the motion of CWAT-01 relative
to the ICM induced through interactions between subclusters 
(SC1 and SC2) and/or interactions between the CWAT-01 parent cluster 
and the other identified clusters.

\subsection { Galaxy cluster assembly }
\label{sec:assembly}

\begin{figure*}
\center{
\includegraphics[bb = 14 14 398 318]{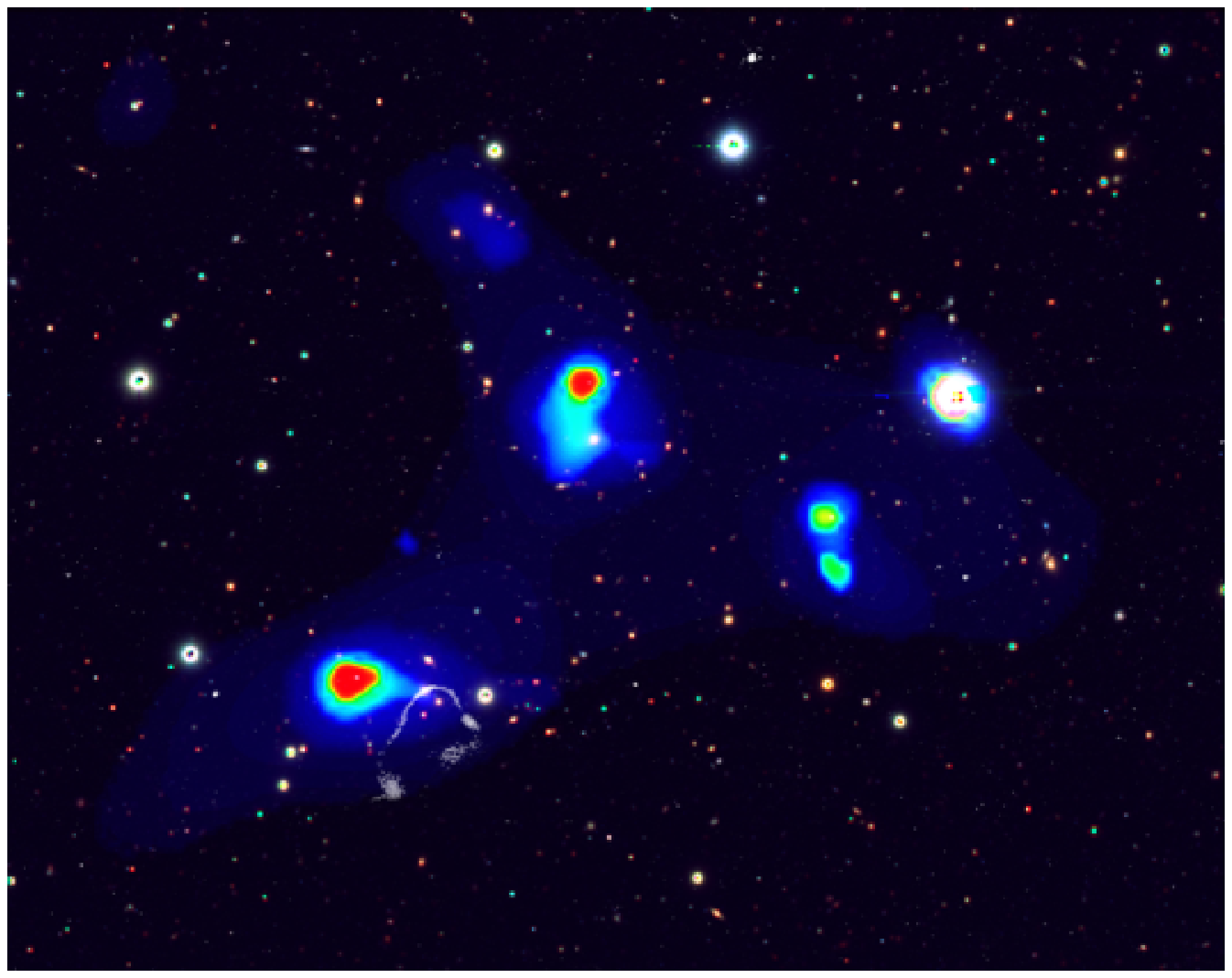}\\
\includegraphics[bb = 14 14 316 236]{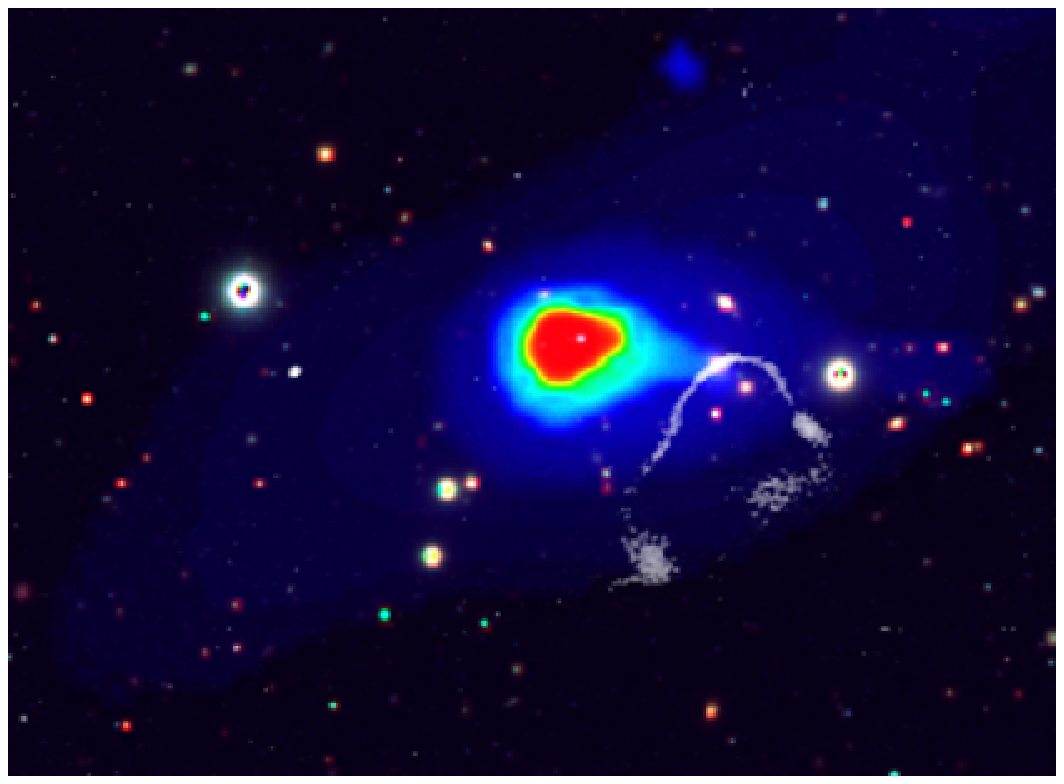}
\caption{ Color composite image of the cluster area (top) and the CWAT-01
  parent cluster (bottom).  The SUBARU $B$ (blue), $V$ (green),
  and $i^+$ (red) bands are displayed in
  the background. The diffuse X-ray emission is presented by rainbow colors and
  the $1.4$~GHz map is shown in white. The top panel encompasses an area of
  $\sim  (2.1   \times 1.7)$~Mpc$^2$  (analogous   to Fig.~\ref{cluster}). The
  size of the bottom panel is as in Fig.~\ref{wat_cluster}. 
\label{color}} }
\end{figure*}

Since CWAT-01 seems to be part of a very complex large-scale structure, the
possible gravitational influence of the other clusters on the galaxy and 
its immediate environment cannot be neglected.
Fig.~\ref{color} shows the distribution of emission from different parts of
the electromagnetic spectrum within the cluster area.
The diffuse X-ray emission has revealed that the CWAT-01 parent cluster
is only one of the poor clusters encompassed in a larger cluster
structure. The cluster assembly contains a minimum of four X-ray
luminous clusters within $\sim 2$ Mpc distance. 
In addition, the VTA indicates that there is at least one more loose
group (LG-N; Fig.~\ref{cluster}) on the outskirts of 
the X-ray cluster assembly but not detected in the X-rays. 
Furthermore, the whole cluster assembly is part of a large-scale structure
component (\lss, Scoville \etal  2006c) extended over $\sim 4$ Mpc 
in NS direction.

 The direction of the jets of CWAT-01 is almost perpendicular to the 
X-ray elongation of the CWAT-01 parent cluster  as well as to the
large-scale elongation of the X-ray emission. Although \lss\ spatially extends
$\sim 4$~Mpc in NS direction, the direction of its {\em long axis}
(as defined in e.g.\ Novikov \etal 1999) is in the NW-SE direction (see
Fig.~3 in Scoville \etal 2006c). Hence, the direction of the jets of CWAT-01
is almost perpendicular also to the long axis of \lss.
This is in
disagreement with the correlation of the alignment between the 
direction of the jets and 
a) the central X-ray emission elongation 
(Burns \etal 1994, Gomez \etal 1997) and 
b) the supercluster axis as defined by the distribution of the nearby Abell
clusters (Novikov \etal 1999). Nevertheless, the misalignment seen 
in CWAT-01's orientation compared to the parent cluster and to \lss~ 
elongations may not be unexpected if we assume an early stage of cluster 
formation. Rich clusters form at the intersection of
large-scale filaments. Compared to numerical simulations 
(e.g.\ Burns \etal 2002, Springel \etal 2005) of cluster  evolution at $z=0.2$ the final cluster has not yet 
formed or relaxed at the intersection of filaments where matter is 
accreted from numerous filaments. 
Thus, in such an environment it would not be unexpected to observe a WAT galaxy
with its jets not aligned with the cluster or large-scale structure
elongations. 

The X-ray and optical analysis indicate that 
each individual poor cluster within the cluster assembly 
is not spherically-symmetric,
both in the diffuse X-ray emission and in the spatial distribution 
of galaxies in the optical (VTA). This strongly indicates that the poor 
clusters are not in a dynamically relaxed state. 
In addition, the overall, large-scale distribution of the cluster assembly is 
very complex and irregular. It is likely that the loose group LG-N is bound to
the system. The lack of X-ray emission from this
group suggests that the system is not very massive.
The flux limit for the cluster search (see \S\ref{sec:Xdata})
allows a detection of an object at $z=0.22$ 
with a limiting luminosity of $L_x^{limit} = 3\times10^{41}$~erg~s$^{-1}$, 
which corresponds to a low-mass group (e.g.\ Mulchaey \etal  2003).
Following the luminosity-temperature (L-T) relation for groups (Osmond \&
Ponman 2004) the limiting temperature then equals $T_x^{limit} \sim
0.4$~keV. Using the mass-temperature (M-T)
relation from Finoguenov \etal (2001) the minimum mass of a system to be
detected in the diffuse $0.5-2$~keV band at $z=0.22$ then corresponds 
to $M_{500}^{tot, limit} \sim 1.5\times 10^{13}M_\odot$.
Thus, the total mass of LG-N has to be less than this mass.

Although spectroscopic verification is needed for the
physical connection of the clusters, we believe that we are witnessing
a protocluster in the process of being built up of multiple galaxy
clusters and groups. 
Adding up the estimated masses for the four clusters identified via 
diffuse X-ray emission (see Tab.~\ref{t:x2}) and the limiting mass
inferred for LG-N the resulting combined mass is
$M\sim 2.0\times 10^{14}\, M_{\odot}$. This is a rough estimate of
the total mass of the cluster system once it is formed since it does not
include the material between the clusters nor other loose groups that may be
bound to the cluster assembly (the cluster assembly is only part of 
\lss). Thus, the estimated mass that the final cluster may have 
after the individual clusters merge would correspond to $\sim 20\%$ of 
the Coma cluster's total mass. 
This is the first time such a complex dynamically young cluster system 
in the process of formation is identified via a WAT radio galaxy.

\section{Summary and Conclusions}
\label{sec:summary}

We have analyzed a wide angle tail (WAT) radio galaxy, CWAT-01, 
first resolved in the VLA-COSMOS survey. 
The multiwavelength data set of the COSMOS survey have 
enabled us to identify and analyze the environment of CWAT-01 in several
independent ways. The cluster structure revealed via CWAT-01 seems to be more
complex than any structure hosting WAT galaxies reported in the past.
We summarize the basic findings of this analysis as follows:
\begin{itemize}
\item 
   The lengths of the Eastern and Western radio jets of CWAT-01 
   are $\sim 210$~kpc and $ \sim 160$~kpc, respectively,  
   and the bending angle is $\sim 100^\circ$ in the projected plane 
   of the sky. 
   It seems to be asymmetric and at this point we cannot rule out  
   that the asymmetry is caused by projection effects.
   The $1.4$ GHz radio power of  
   $P_{1.4} = 2.0\times 10^{24} \,\mathrm{W\,Hz^{-1}}$
   puts CWAT-01 on the lower end of the FRI-II break 
   region where WATs are usually found.
\item  
   The host galaxy of CWAT-01 is an elliptical galaxy with a shallower
   surface brightness profile than predicted by the de Vaucouleurs law.
   It is the brightest cluster galaxy (BCG) in the
   CWAT-01 parent cluster. The surface brightness profile is 
   very well fit by the Sersic $r^{1/n}$ law with 
   $n=5$, $r_{\mathrm{eff}}=8.2$~kpc and 
   $\mu_{\mathrm{eff}}=22.11 \, \mathrm{mag\, arcsec^{-2}}$ 
   consistent with 
   values typical for  brightest cluster galaxies (cD/D).
\item 
   Applying several hydrodynamical models, taking ram pressure and buoyancy 
   forces into account, to explain the observed
   bending of the radio jets of CWAT-01, the allowed range of the galaxy 
   velocity relative to the ICM is approximately 
   $300-550$\kms. Both, the upper and lower velocity are higher
   than is expected for dominant galaxies (i.e. BCGs) in relaxed 
   systems.
\item 
   The cluster hosting CWAT-01 (CWAT-01 parent cluster) was detected 
   in diffuse X-ray emission. The luminosity 
   weighted temperature of the cluster is $\sim 1.7$~keV 
   consistent with poor cluster temperatures. 
   The total mass within the $r_{500}$ radius is 
   $\sim 5.8\times 10^{13}\, M_\odot$. The cluster shows
   evidence for subclustering, both in diffuse X-ray emission and in the
   spatial distribution of galaxies found from the optical analysis
   applying the Voronoi tessellation-based approach  (VTA). The
   distribution of early type galaxies is not centrally concentrated, it 
   is irregular and partitioned into two apparently distinct subclumps.
\item The CWAT-01 parent cluster itself is part of a larger cluster assembly
   consisting of a minimum of 4 clusters within $\sim 2$~Mpc distance
   identified via diffuse  X-ray emission. The ICM temperatures of the three
   clusters surrounding the CWAT-01 parent cluster
   are in the $1.4-1.5$~keV range consistent with temperatures of poor
   clusters. 
   The total masses of the clusters within the $r_{500}$ radius are  
   in the range of about $(4.1-4.4) \times 10^{13} M_\odot$. 
\item The Voronoi tessellation-based approach (VTA) results indicate that there
   is at least one more loose group that 
   is likely  bound to the system. From the X-ray detection limit for
   diffuse sources we infer that the total mass of this group
   must be less than $ 1.5\times 10^{13}M_\odot$.

\end{itemize}

The whole cluster structure described in this paper is encompassed 
in a large-scale structure component, \lss, reported in
Scoville \etal  (2006c). \lss~is elongated in NS direction 
and extends $\sim 4$ Mpc along the major axis.
Our results strongly indicate that we are
witnessing the formation of a large cluster from an assembly of 
multiple clusters, consistent with the scenario of hierarchical structure 
formation. If this is the case, then the estimated minimum total mass of the 
final single cluster after the poor clusters merge would correspond to 
$M\sim 2.0\times 10^{14}\, M_{\odot}$ or $\sim20\%$ of the Coma cluster mass. 
In this scenario, the large velocity of the CWAT-01 host galaxy relative 
to the ICM can easily be explained. The CWAT-01 parent cluster
seems not to be relaxed, thus a plausible explanation of the motion of
the galaxy relative to the ICM is interaction of the two identified 
subclumps (SC1 and SC2) within the cluster. 
On the other hand, we cannot rule out the 
gravitational influence of the other poor clusters as a cause for 
inducing such a velocity. 

Resolving the detailed physics causing the bending of the radio jets of
CWAT-01, the dynamical interplay between and 
within particular clusters as well as the spectroscopic confirmation 
of the physical connection of the clusters has to await the completion of the 
zCOSMOS program (Lilly \etal 2006). 
 Nevertheless, our results support the idea that WAT galaxies are
  tracers of galaxy clusters, in particular dynamically young ones.

\acknowledgments 

VS thanks Rachel Somerville and Klaus Meisenheimer for insightful discussions.
CC, ES and VS acknowledge support from NASA grant HST-GO-09822.31-A. 
CC would like to  acknowledge support from the
Max-Planck Society and the Alexander von Humboldt Foundation through
the Max-Planck-Forschungspreis 2005.
AF acknowledges support from BMBF/DLR under
grant 50 OR 0207, MPG and a partial support from NASA grant NNG04GF686.
IS acknowledges the support of the European Community under a
Marie Curie Intra-European Fellowship.
CSB ackgnowledges funding by the Deutsche Forschungsgemeinschaft.
KJ acknowledges support by the German DFG under
grant SCHI 536/3-1.

The XMM-Newton project is an ESA Science Mission with instruments
and contributions directly funded by ESA Member States and the
USA (NASA). The XMM-Newton project is supported by the Bundesministerium f\"ur
Wirtschaft und Technologie/Deutsches Zentrum f\"ur Luft- und Raumfahrt
(BMWI/DLR, FKZ 50 OX 0001), the Max-Planck Society and the
Heidenhain-Stiftung, and also by PPARC, CEA, CNES, and ASI.  Part of this work
was also supported by the Deutsches Zentrum f\"ur Luft-- und Raumfahrt, DLR
project numbers 50 OR 0207 and 50 OR 0405. 

    Funding for the Sloan Digital Sky Survey (SDSS) has been provided by the
    Alfred P. Sloan Foundation, the Participating Institutions, the National
    Aeronautics and Space Administration, the National Science Foundation, the
    U.S. Department of Energy, the Japanese Monbukagakusho, and the Max Planck
    Society. The SDSS Web site is http://www.sdss.org/. 

    The SDSS is managed by the Astrophysical Research Consortium (ARC) for the
    Participating Institutions. The Participating Institutions are The
    University of Chicago, Fermilab, the Institute for Advanced Study, the
    Japan Participation Group, The Johns Hopkins University, the Korean
    Scientist Group, Los Alamos National Laboratory, the Max-Planck-Institute
    for Astronomy (MPIA), the Max-Planck-Institute for Astrophysics (MPA), New
    Mexico State University, University of Pittsburgh, University of
    Portsmouth, Princeton University, the United States Naval Observatory, and
    the University of Washington.




\clearpage

\end{document}